\documentclass[a4paper,onecolumn,usenatbib]{mnras}

\usepackage{newtxtext,newtxmath}

\usepackage{hyperref}
\usepackage[T1]{fontenc}
\usepackage{ae,aecompl}
\usepackage{longtable} 
\usepackage{hhline}

\usepackage{graphicx}	
\usepackage{amsmath}	
\usepackage{amssymb}	

\title[MFDMA of \textit{Kepler} stars with surface differential rotation traces]{Multifractal detrended moving average analysis of \textit{Kepler} stars with surface differential rotation traces}

\author[D. B. de Freitas et al.]{
D. B. de Freitas,$^{1}$\thanks{E-mail: danielbrito@fisica.ufc.br (DBdeF)} M. M. F. Nepomuceno,$^{2}$
J. G. Cordeiro,$^{1}$ M. L. Das Chagas,$^{3}$ \newauthor and J. R. De Medeiros$^{4}$
\\
$^{1}$Departamento de F\'{\i}sica, Universidade Federal do Cear\'a, Caixa Postal 6030, Campus do Pici, 60455-900 Fortaleza, Cear\'a, Brazil\\
$^{2}$Departamento de Ci\^encia e Tecnologia, Universidade Federal Rural do Semi-\'Arido, Campus Cara\'ubas, Rio Grande do Norte, Brazil\\
$^{3}$Faculdade de F\'{\i}sica - Instituto de Ci\^encias Exatas, Universidade Federal do Sul e Sudeste do Par\'a, Marab\'a, PA 68505-080, Brazil\\
$^{4}$Departamento de F\'{\i}sica, Universidade Federal do Rio Grande do Norte, 59072-970 Natal, RN, Brazil}

\date{Accepted XXX. Received YYY; in original form ZZZ}

\pubyear{2015}

\begin{document}
\label{firstpage}
\pagerange{\pageref{firstpage}--\pageref{lastpage}}
\maketitle

\begin{abstract}
A multifractal formalism is employed to analyse high-precision time-series data of \textit{Kepler} stars with surface differential rotation traces. The multifractal detrended moving average analysis (MFDMA) algorithm has been explored to characterize the multi-scale behaviour of the observed time series from a sample of 662 stars selected with parameters close to those of the Sun, e.g., effective temperature, mass, effective gravity and rotation period. Among these stars, 141 have surface differential rotation traces, whereas 521 have no detected differential rotation signatures. In our sample, we also include the Sun in its active phase. Our results can be summarized in two points: first, our work suggests that starspots for time series with and without differential rotation have distinct dynamics, and second, the magnetic fields of active stars are apparently governed by two mechanisms with different levels of complexity for fluctuations. Throughout the course of the study, we identified an overall trend whereby the differential rotation is distributed in two $H$ regimes segregated by the degree of asymmetry $A$, where $H$-index denotes the global Hurst exponent which is used as a measure of long-term memory of time series. As a result, we show that the degree of asymmetry can be considered a segregation factor that distinguishes the differential rotation behaviour when related to the effect of the rotational modulation on the time series. In summary, the multifractality signals in our sample are the result of magnetic activity control mechanisms leading to activity-related long-term persistent signatures.
\end{abstract}

\begin{keywords}
stars: rotation -- methods: data analysis -- stars: solar-type
\end{keywords}



\section{Introduction}

Differential rotation is a key parameter for understanding the stellar dynamo and transport processes in the interiors of stars. Different methods can be used to measure differential rotation (DR). For example, the rotational broadening of spectral line profiles is directly measurable by means of deconvolution techniques that are based on Fourier analysis \citep[e.g.,][]{Reiners2003} or Doppler imaging and Zeeman-Doppler imaging \citep[e.g.,][]{Strassmeier2009}. Recently, \cite{lanza2014} showed that the lifetime of the starspot pattern can be measured from the decay of the height of successive autocorrelation peaks and that the lifetime determined in this manner affects the possibility of measuring the DR with a two-spot model.

The \textit{Kepler} space mission now offers the ability to measure the rotation periods of thousands of stars based on high-precision photometry \citep{boru}. This measurement opens a new perspective for the study of DR. In a recent paper, \cite{daschagas2016} identified, from the entire \textit{Kepler} database of time series for more than 800 stars observed over 17 quarters, 17 stars with the signature of DR and with sufficiently stable signals. The authors used a simple two-spot model combined with a Bayesian information criterion for this sample in the search to measure the amplitude of surface DR \citep{lanza2014}. Prior to that paper, \cite{reinhold} used a procedure based on the Lomb-Scargle periodogram for a pre-whitening approach, resulting in a large sample of stars with the DR signature.

A substantial portion of the database from \cite{daschagas2016} belongs to the \cite{reinhold} sample. In that sample, they extended the number to over 40,000 active stars, for which 24,124 primary rotation periods ($P1$) between 0.5 and 45 days were found, with a mean of $\langle P1 \rangle$=16.3 days. In addition, the authors found evidence of surface DR in 18,616 stars.

In the present study, we analysed the multifractal nature of a sample of 662 active stars extracted from \cite{reinhold} using the multifractal detrending moving average (MFDMA) algorithm, which was proposed by \cite{gu2010} and tested by \cite{defreitas2016,defreitas2017} for \textit{Kepler} and CoRoT stars.

In general, multifractal analysis and its different methods and procedures \citep{Kantelhardt,gu2010,tang}, which were developed over more than 5 decades, are applied in a wide variety of fields as inspired by \cite{hurst1951,mw1969a,mw1969b,feder1988}. In several areas, such as medicine \citep{ivanov1999} and geophysics \citep{telesca2006,db2008,defreitas2013b}, multifractality has already been adopted as a determinant approach for analysing the behaviours of time series with nonlinearity, nonstationarity and correlated noise, which are just a few of the properties that this analysis is able to describe \citep{movahed,Norouzzadeha,sps2009,seuront,a2011}. More recently, \cite{droz} showed that the multifractal analysis of sunspot numbers is a crucial procedure for understanding the behaviour of the magnetic field of the Sun.

Beginning in 2013, \cite{defreitas2013,defreitas2016,defreitas2017} showed that multifractality analysis of stellar activity is a powerful tool for estimating physical properties based on the multifractality spectrum. More specifically, a set of four multifractal indexes that are extracted from geometric features of the singularity spectrum (see the next section) are used to describe the fluctuations in the different scales. In particular, these authors showed that the Hurst exponent has a strong correlation with the rotation period when used for a wide range of periods. In the present work, we restricted the interval of the period to a solar regime of between 24.5 and 33.5 days. Our motivation was to eliminate the correlation cited above and, therefore, verify whether there are other multifractal indexes affected by stellar rotation. This procedure can be useful for investigating details in a time series that a conventional method cannot determine.

As mentioned by \cite{elia}, multifractal analysis is an under-exploited approach for quantifying and investigating the behaviour of astrophysical time series that exhibit magnetic activity. Indeed, most papers \citep[e.g.,][]{affer,demedeiros2013,mathur2014a,aigrain} that address time series still use procedures that summarize the photometry of the star as a purely Gaussian (normal) process and use methods that rely on a statistically limited analysis. Among the limitations are the lack of criteria for analysing the presence of memory with noisy backgrounds and the shapes of the probability distributions, which are associated with the dependences of different statistical moments. These canonical methods consider only the first (i.e., mean) and second (i.e., variance) statistical moments. Thus, the analysis omits other moments, such as asymmetry, kurtosis and higher-order moments, which are denoted by parameter $q$ in the multifractal scenario (see the next section). However, multifractal time series are not normally distributed, and all $q$-order statistical moments should be considered \citep{Kantelhardt,ihlen}.

Our paper is organized as follows. In Section 2, we provide a detailed description of multifractal analysis in which we emphasize a set of indexes that are extracted from the multifractal spectrum. In Section 3, we describe the working sample. Section 4 is dedicated to a detailed discussion of the results. In the last section, our final remarks are summarized.

\begin{figure*}
	\includegraphics[width=1.0\columnwidth]{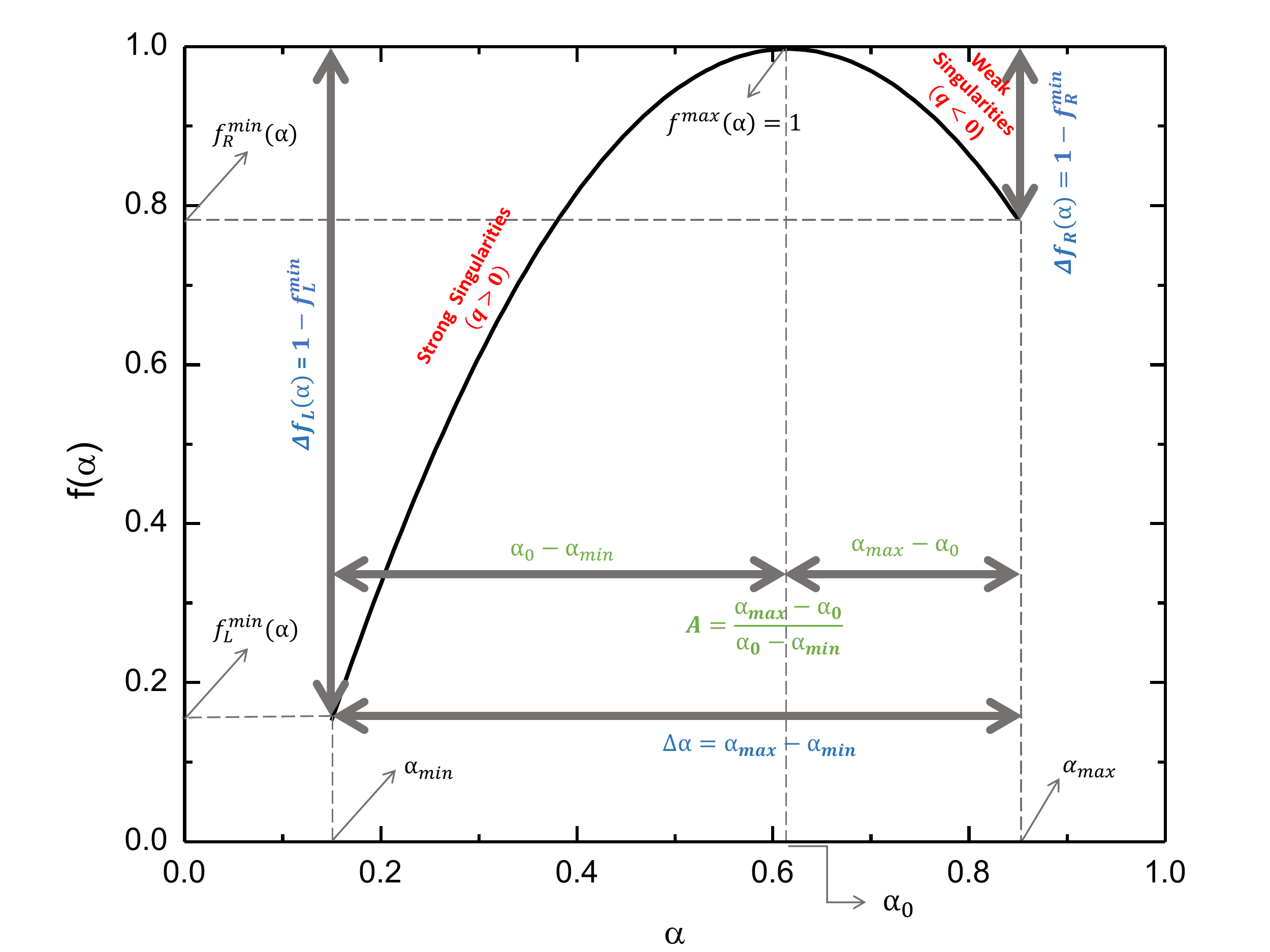}
	\caption{Schematic picture of the multifractal spectrum descriptors with the use of geometric language.}
	\label{figMFDMA}
\end{figure*}

\begin{figure*}
	\includegraphics[width=1.0\columnwidth]{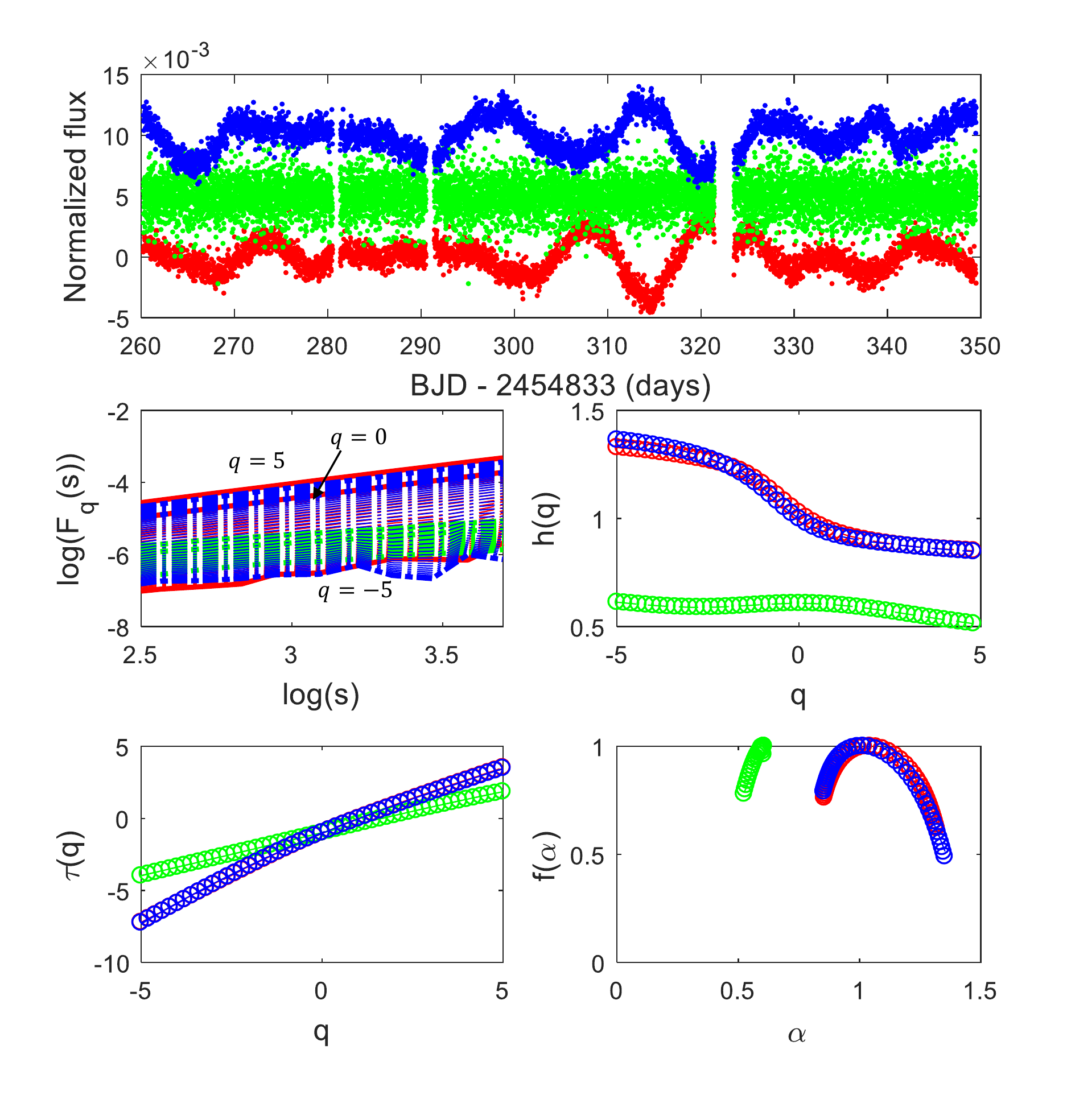}
	\caption{Multifractal analysis of normalized PDCSAP flux for KIC002450531 star following steps 5 and 6 (see Section 2). This star presents a primary rotation period of 22.5 days and a secondary rotation period of 18.2 days. \textit{Top panel}: Time series of KIC002450531 (in red), a star with DR traces (see Section 3 and Table 1). The shuffled (green) and surrogate (blue) time series are based on the procedure mentioned in Section 2.2. \textit{Left middle}: the multifractal fluctuation function $F_{q}(s)$ obtained from the MFDMA method. Each curve corresponds to a different fixed value of $q=-5,...,5$ (with a step of 0.2) from bottom to top, where red lines are original data, green lines are shuffled data and blue lines are surrogate data. \textit{Right middle}: $q$-order Hurst exponent ($h(q)$) as a function of $q$-parameter. This panel shows the truncation originated from the levelling of $h(q)$ for positive $q$ values. This truncation plays an important role in the measurement of $\Delta f_{L}(\alpha)$. \textit{Left bottom panel}: comparison of the multifractal scaling exponent $\tau(q)$ of the original (red), shuffled (green) and surrogate (blue) time series. In this panel, it is possible to identify a crossover in $q=0$, a typical feature of the multifractal time series. \textit{Right bottom panel}: multifractal spectra $f(\alpha)$ of the original (red), shuffled (green) and surrogate (blue) time series.}
	\label{fig1a}
\end{figure*}

\begin{figure*}
	\includegraphics[width=1.0\columnwidth]{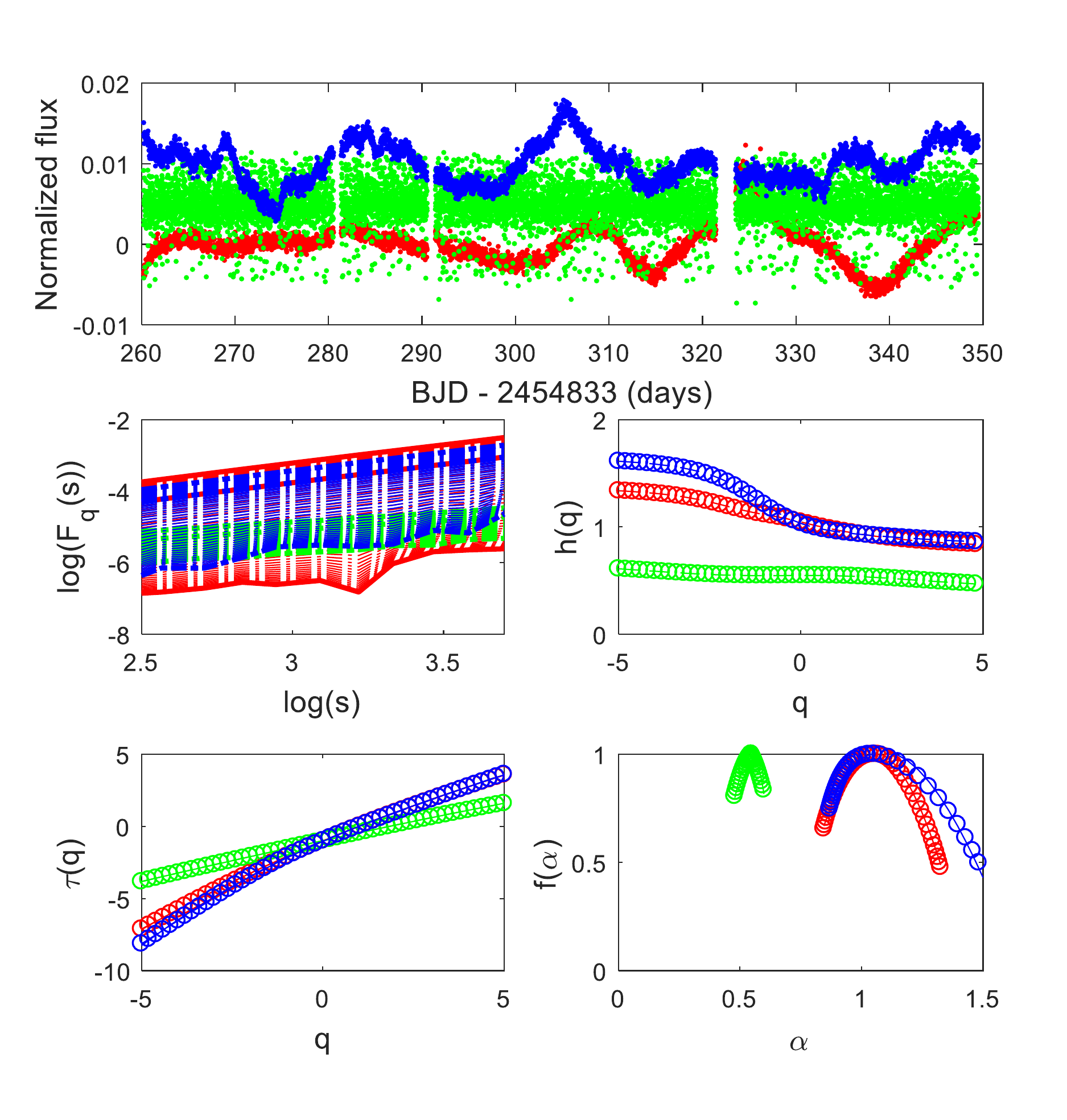}
	\caption{Multifractal analysis of normalized SAP4 flux for KIC002450531 star following steps 5 and 6 (see Section 2).}
	\label{fig1a2}
\end{figure*}

\begin{figure*}
	\includegraphics[width=1.0\columnwidth]{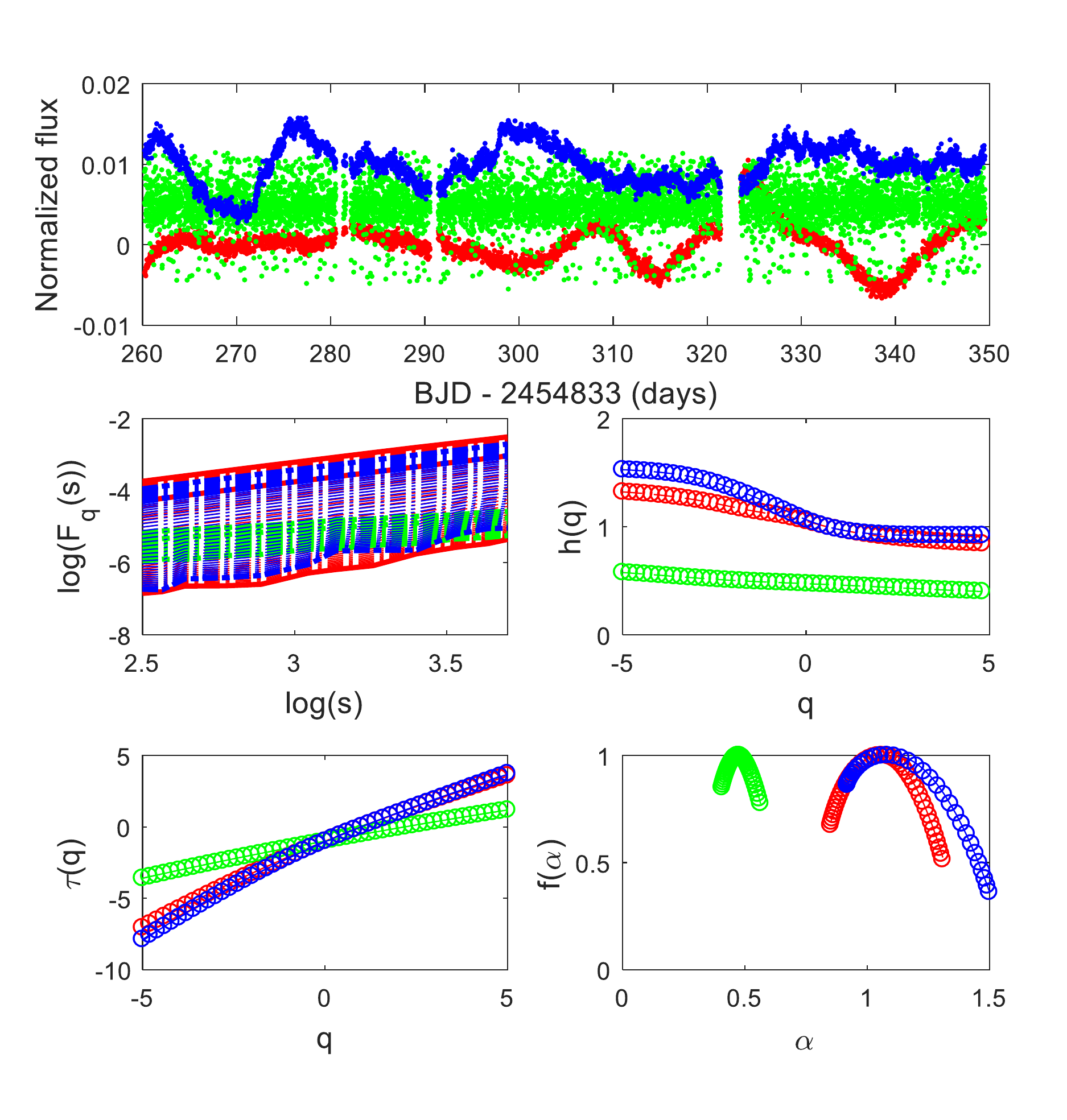}
	\caption{Multifractal analysis of normalized SAP4+QUALITY flux for KIC002450531 star following steps 5 and 6 (see Section 2).}
	\label{fig1a3}
\end{figure*}

\begin{figure*}
	\includegraphics[width=1.0\columnwidth]{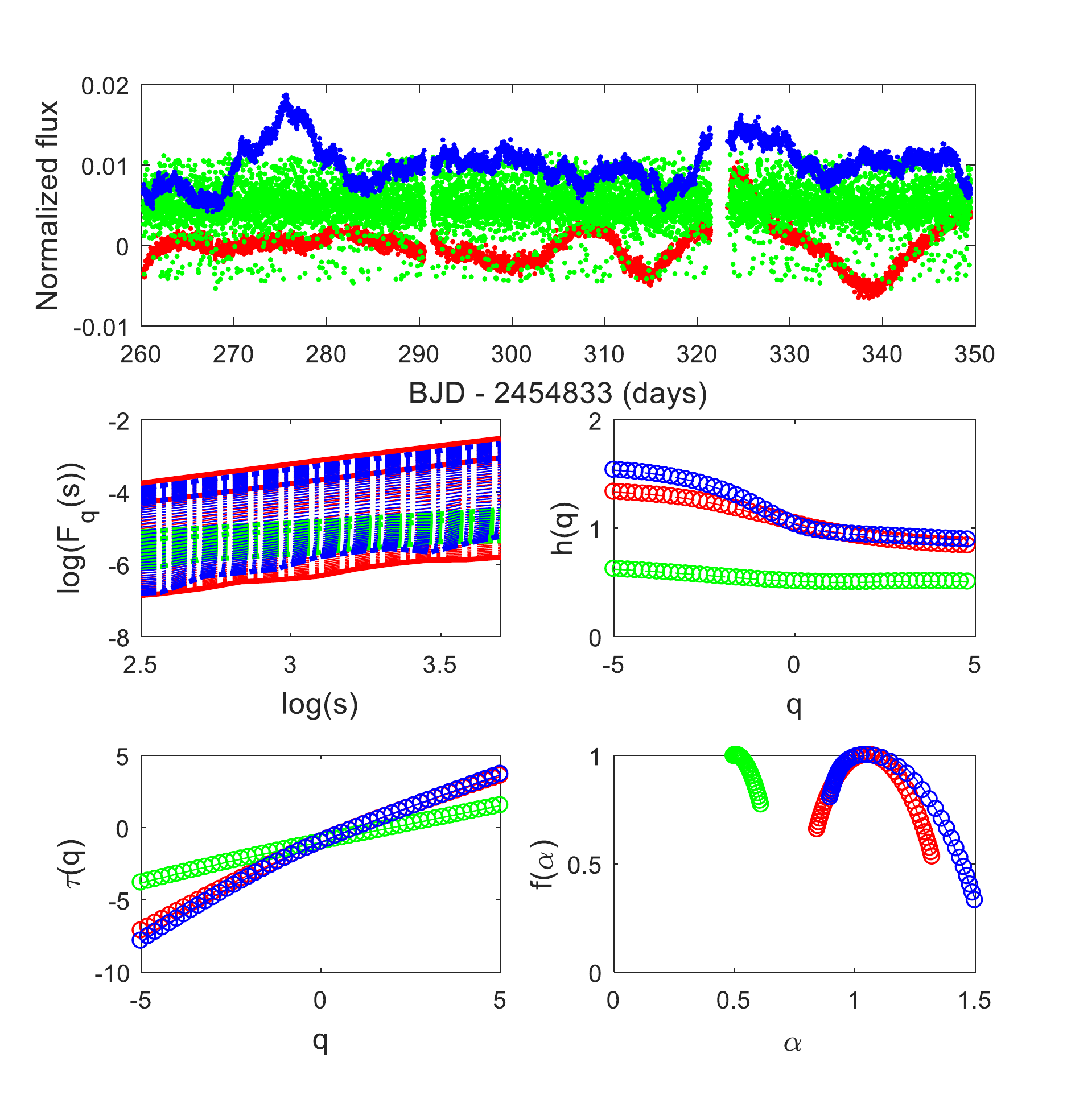}
	\caption{Multifractal analysis of normalized SAP4+QUALITY+LINEAR INTERPOLATION flux for KIC002450531 star following steps 5 and 6 (see Section 2).}
	\label{fig1a4}
\end{figure*}

\begin{figure*}
	\includegraphics[width=1.0\columnwidth]{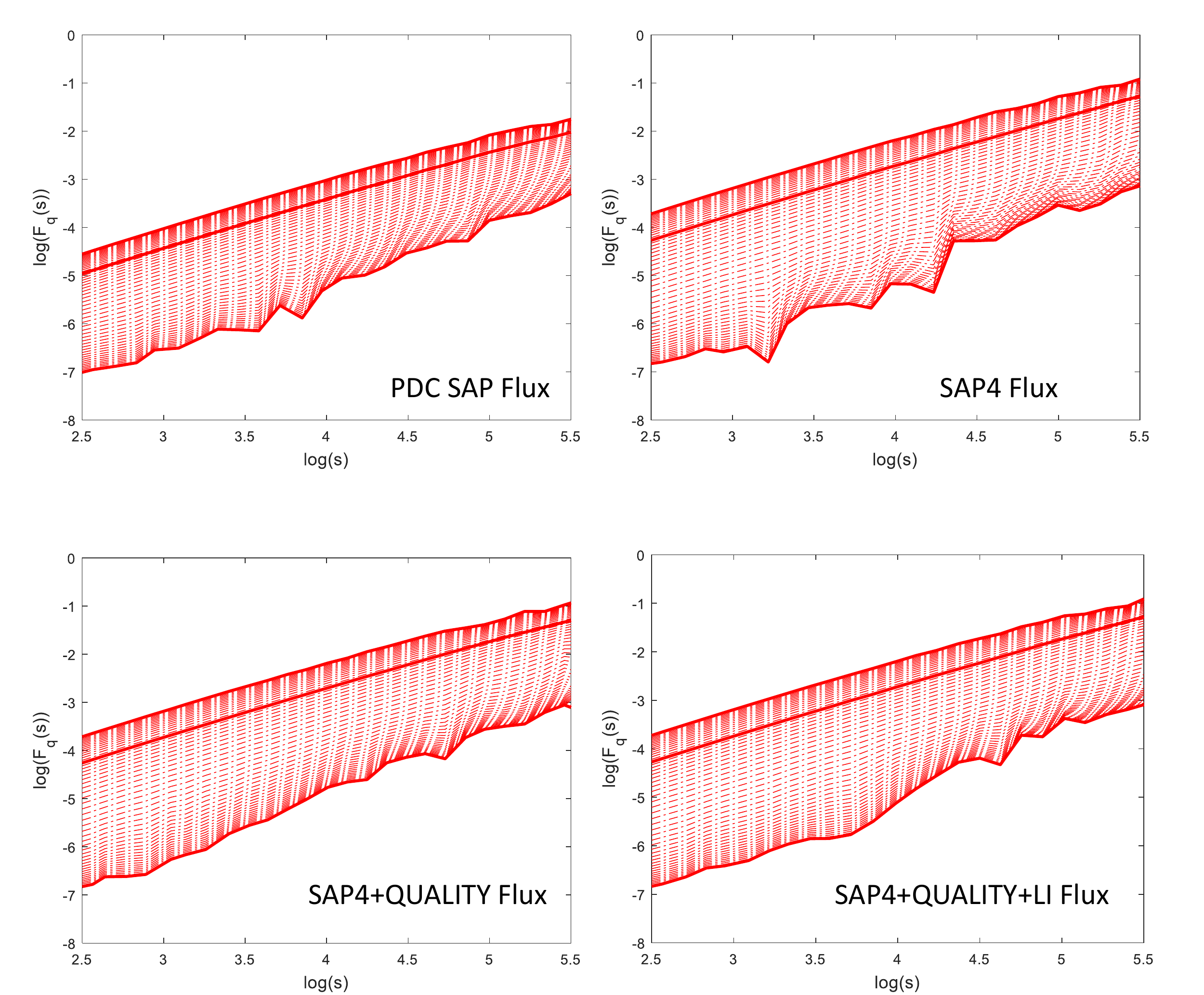}
	\caption{Log–log plot for $F_{s}(q)$ vs. scale $s$ obtained from MFDMA for original KIC002450531 time series using different types of data as mentioned by Figs. from \ref{fig1a} to \ref{fig1a4}.}
	\label{figFsq}
\end{figure*}

\begin{figure*}
	\includegraphics[width=1.0\columnwidth]{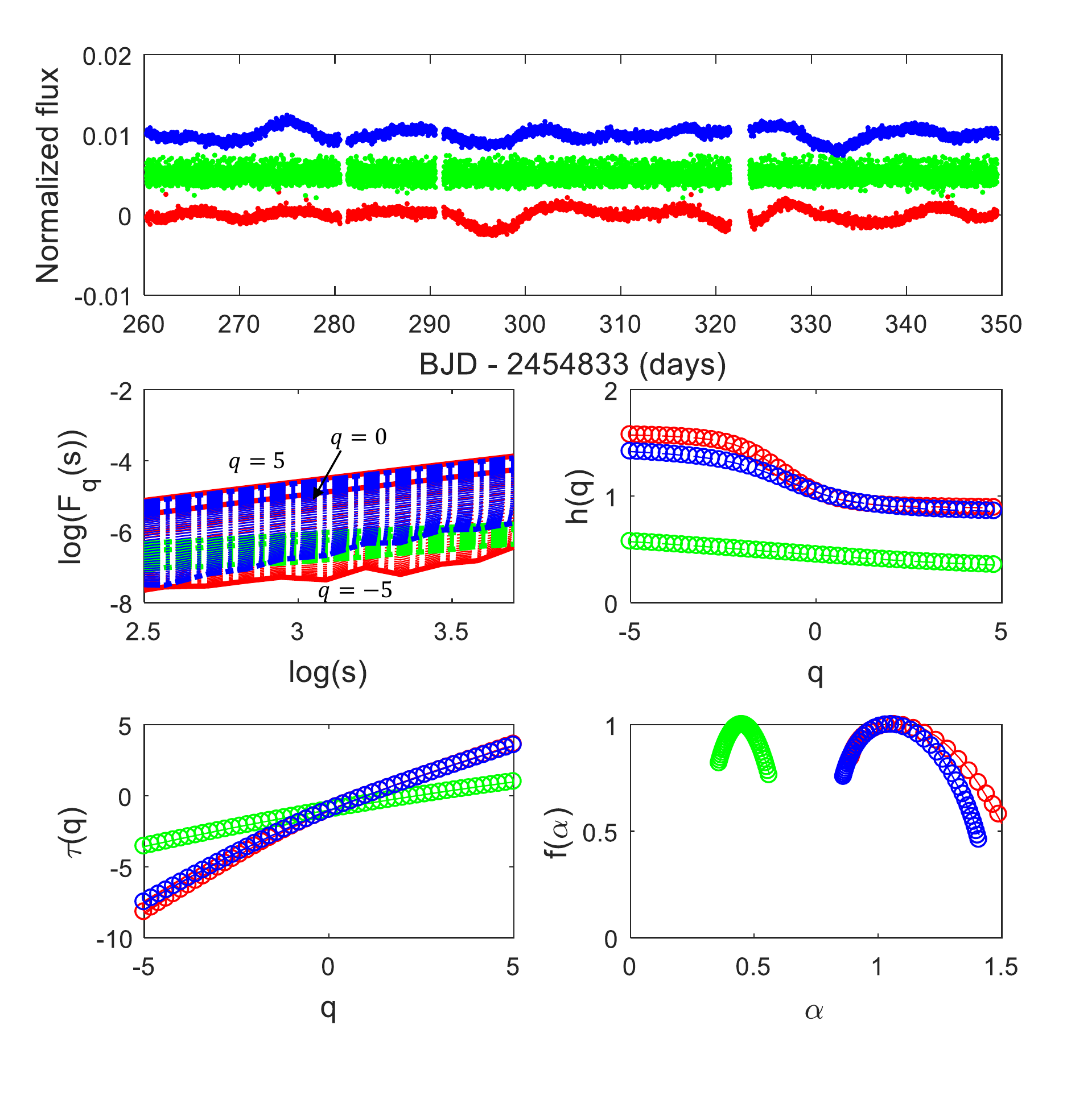}
	\caption{Multifractal analysis of normalized PDCSAP flux for the Kepler star named KIC009963569, a star without DR traces. This star presents a rotation period of 26.1 days.}
	\label{fig1b}
\end{figure*}

\begin{figure*}
	\includegraphics[width=1.0\columnwidth]{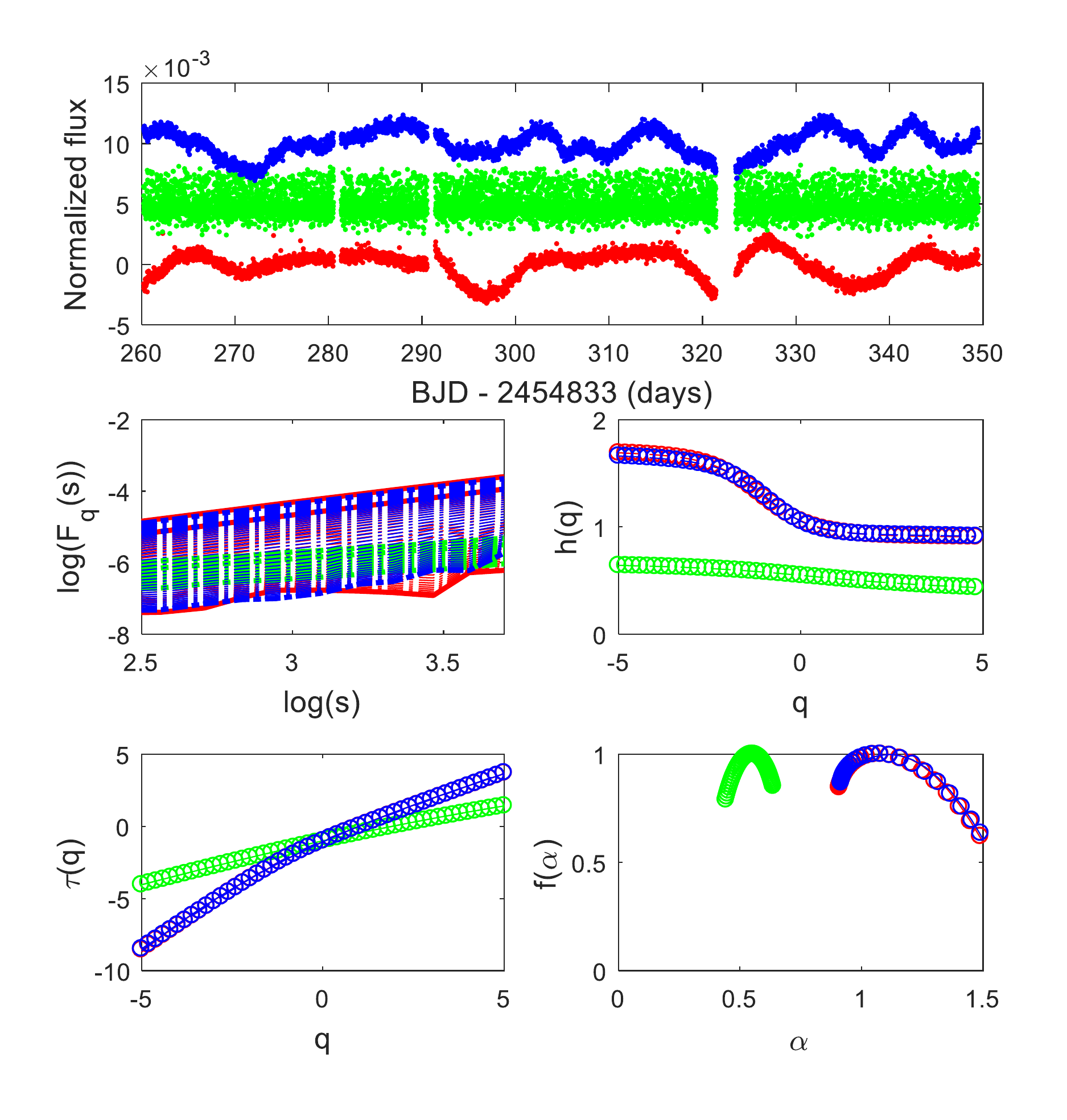}
	\caption{Multifractal analysis of normalized SAP4 flux for KIC009963569.}
	\label{fig1c}
\end{figure*}

\begin{figure*}
	\includegraphics[width=1.0\columnwidth]{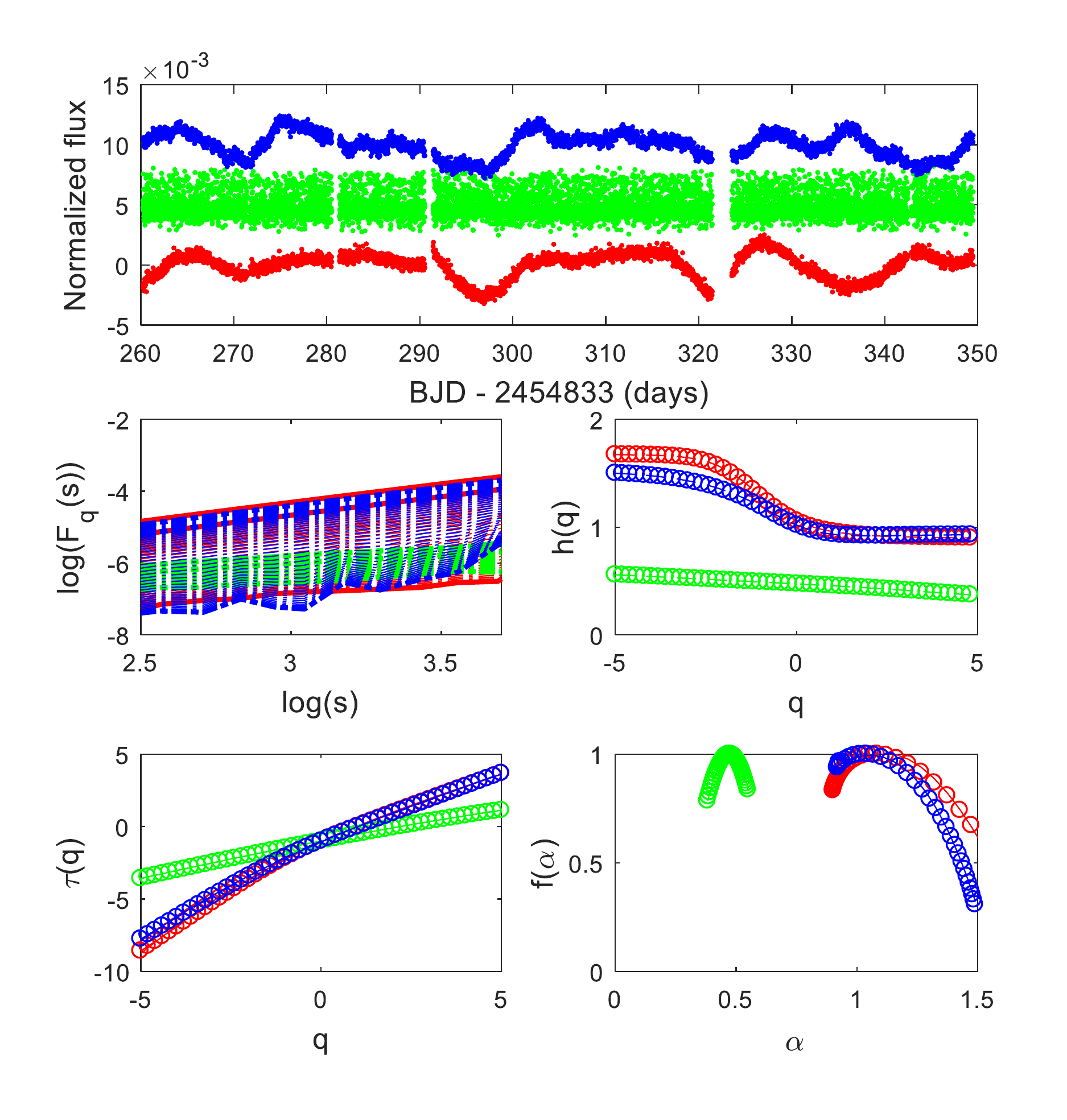}
	\caption{Multifractal analysis of normalized SAP4+QUALITY flux for KIC009963569.}
	\label{fig1d}
\end{figure*}

\begin{figure*}
	\includegraphics[width=1.0\columnwidth]{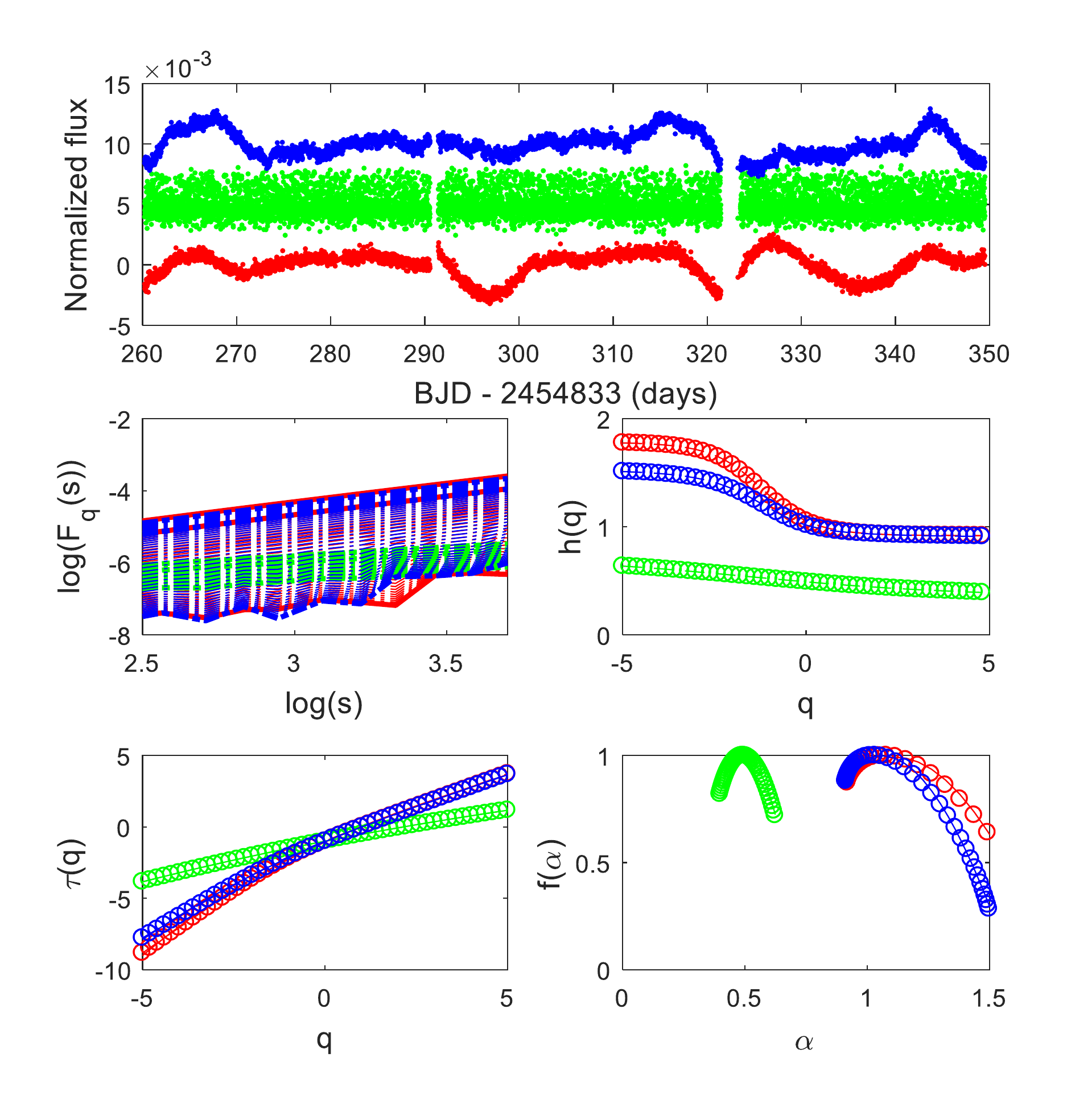}
	\caption{Multifractal analysis of normalized SAP4+QUALITY+LINEAR INTERPOLATION flux for KIC009963569.}
	\label{fig1e}
\end{figure*}

\begin{figure*}
	\includegraphics[width=1.0\columnwidth]{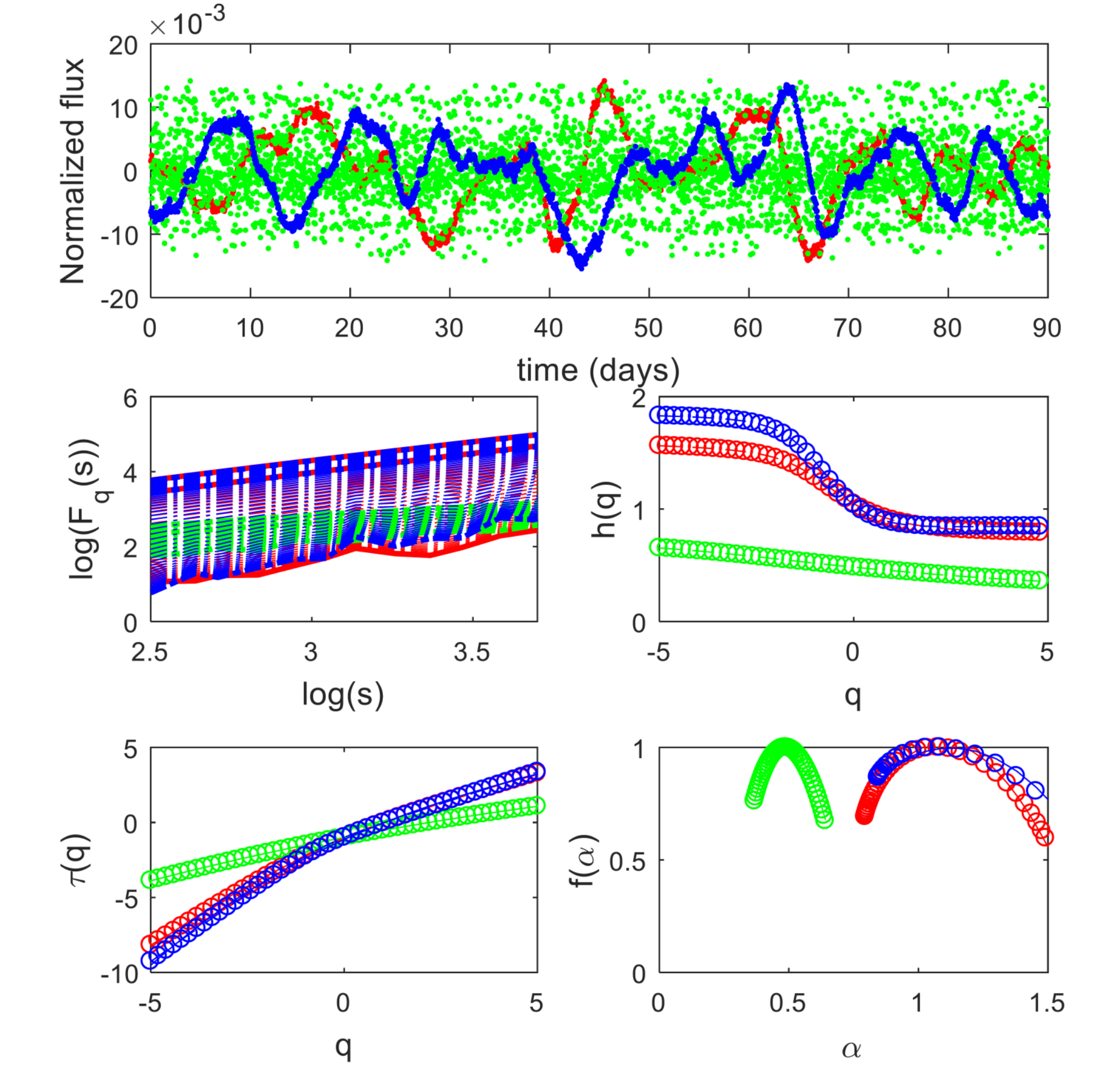}
	\caption{Multifractal analysis of the normalized VIRGO/SPM (Green + Red channels) instrument obtained as described in Section 3. For the Sun, the four multifractal parameters obtained are given by $A=1.72$, $\Delta\alpha=0.76$, $\Delta f_{L}=0.31$, $\Delta f_{R}=0.60$ and $H=0.84$.}
	\label{figsolar}
\end{figure*}

\section{Multifractal analysis}
Recently, \cite{defreitas2017} used multifractal analysis to investigate the multi-scale behaviour of a set of 34 M dwarf stars observed by the Kepler mission as well as the Sun in its active phase. \cite{defreitas2016,defreitas2017} showed that the MFDMA algorithm, which was developed by \cite{gu2010}\footnote{MATLAB codes for MFDMA analysis can be found in the \texttt{arXiv} version of \cite{gu2010}'s paper: \texttt{https://arxiv.org/pdf/1005.0877v2.pdf}} and \cite{tang}, is a powerful technique that provides invaluable information on the dynamic structure of a time series.

\cite{gu2010} summarized the MFDMA algorithm in the following steps:

\begin{itemize}
	\item Step 1: Calculate the time-series profile over time $t=1,2,3,...,N$:
\end{itemize}
\begin{equation}
\label{eq1}
y(t)=\sum^{t}_{i=1}x(i), \quad t=1,2,3,...,N,
\end{equation}
where the above equation is a sequence of cumulative sums and $x(i)$ is the observed time series.

\begin{itemize}
	\item Step 2: Calculate the moving average function of Eq. (\ref{eq1}) in a moving window:
\end{itemize}
\begin{equation}
\label{eq1a}
\tilde{y}(t)=\frac{1}{s}\sum^{\left\lceil s-1\right\rceil}_{k=0}y(t-k),
\end{equation}
where $s$ is the window size, and $\left\lceil (x)\right\rceil$ is the smallest integer that is not smaller than argument $(x)$.

\begin{itemize}
	\item Step 3: Detrend the series by removing the moving average function, $\tilde{y}(i)$, and obtain the residual sequence, $\epsilon(i)$, through:
\end{itemize}
\begin{equation}
\label{eq2}
\epsilon(i)=y(i)-\tilde{y}(i).
\end{equation}
where $s \le i\le N$. The residual series, $\epsilon(i)$, is divided into $N_{s}$ disjoint segments with the same size of $s$. In our code, $N_{s}$ is an input parameter, which we set to a value of 30, as suggested by \cite{gu2010}. On the other hand, the box size $s$ is an output parameter and varies in the range of $[10,N/10]$. In addition, each segment can be expressed by $\epsilon_{\nu}$, where $\epsilon_{\nu}(i)=\epsilon(l+i)$ such that $1\le i \le s$ and $l=(\nu-1)s$.

\begin{itemize}
	\item Step 4: Calculate the root-mean-square (RMS) fluctuation function, $F_{\nu}(s)$, for a segment of size $s$:
\end{itemize}
\begin{equation}
\label{eq3}
F_{\nu}(s)=\left\{\frac{1}{s}\sum^{s}_{i=1}\epsilon^{2}_{\nu}(i)\right\}^{\frac{1}{2}}.
\end{equation}

\begin{itemize}
	\item Step 5: Generate the function, $F_{q}(s)$, of the $q$th order:
\end{itemize}
\begin{equation}
	\label{eq4}
	F_{q}(s)=\left\{\frac{1}{N_{s}}\sum^{N_{s}}_{\nu=1}F^{q}_{\nu}(s)\right\}^{\frac{1}{q}},
	\end{equation}
for all $q\neq 0$. According to the \cite{gu2010}’s formalism, we assume $q$ to be a real number in the interval $[-5,5]$. The $q$th-order function is the statistical moment (\textit{e.g.}, for $q$=2, we have the variance), and for $q=0$,
\begin{equation}
	\label{eq4b}
	\ln\left[F_{0}(s)\right]=\frac{1}{N_{s}}\sum^{N_{s}}_{\nu=1}\ln [F_{\nu}(s)],
	\end{equation}
where the scaling behaviour of $F_{q}(s)$ follows the relationship that is given by $F_{q}(s)\sim s^{h(q)}$, and $h(q)$ denotes the Holder exponent or generalized Hurst exponent. Each value of $q$ yields a slope, $h$.

\begin{itemize}
	\item Step 6: Given $h(q)$, the multifractal scaling exponent, $\tau(q)$, can be computed:
\end{itemize}
\begin{equation}
\label{eq5}
\tau(q)=q h(q)-1.
\end{equation}

Finally, the singularity strength function, $\alpha(q)$, and the multifractal spectrum, $f(\alpha)$, are obtained via a Legendre transform:
\begin{equation}
\label{eq7}
\alpha(q)=\frac{d\tau(q)}{dq}
\end{equation}
and
\begin{equation}
\label{eq6}
f(\alpha)=q\alpha-\tau(q).
\end{equation}
In addition, for a monofractal signal, $h$ is the same for all values of $q$. For a multifractal signal, $h(q)$ is a function of $q$, and the multifractal spectrum is parabolic (see Figure \ref{figMFDMA}). In particular, for $q=2$, we have a specific value of $h$ denoted by $H$ and known as the global Hurst exponent.

\subsection{Multifractal indexes}
We tested the set of four multifractal descriptors that were extracted from the spectrum $f(\alpha)$, as proposed by \cite{defreitas2017}. A scheme of indexes used to quantify the multifractal spectrum in this work is shown in Figure \ref{figMFDMA}. \cite{defreitas2017} have used this same figure to describe the shape of the multifractal spectrum. Several types of scaling parameters are used to parameterize the multifractal structure of the time series. Here, we list four parameters, which were identified by \cite{defreitas2017}:

\subsubsection{Degree of asymmetry ($A$)}
This index, which is also called the skewness in the shape of the $f(\alpha)$ spectrum, is expressed as the following ratio:
\begin{equation}
\label{eq8}
A=\frac{\alpha_{max}-\alpha_{0}}{\alpha_{0}-\alpha_{min}},
\end{equation}
where $\alpha_{0}$ is the value of $\alpha$ when $f(\alpha)$ is maximal. The value of this index $A$ indicates one of three shapes: right-skewed ($A>1$), left-skewed ($0<A<1$) or symmetric ($A=1$). The right endpoint $\alpha_{max}$ and the left endpoint $\alpha_{min}$ represent the maximum and minimum values of the singularity strength function, $\alpha(q)$, respectively.

\subsubsection{Degree of multifractality ($\Delta \alpha$)}
This index represents the broadness:
\begin{equation}
\label{eq9}
\Delta \alpha=\alpha_{max}-\alpha_{min},
\end{equation}
where $\alpha_{max}$ and $\alpha_{min}$ are as defined above. A low value of $\Delta\alpha$ indicates that the time series is close to fractal, and the multifractal strength increases when $\Delta\alpha$ increases \citep{defreitas2009,defreitas2017}.

\subsubsection{The singularity parameters $\Delta f_{L}$, $\Delta f_{R}$ and $C$}
The parameters $\Delta f_{L}(\alpha)$ and $\Delta f_{R}(\alpha)$ characterize the broadness, which is defined as the difference between the maximum $(f^{max}(\alpha)=1)$ and minimum values of the singularity spectrum, where the left-side endpoint is denoted by $f^{min}_{L}(\alpha)$ and the right-side endpoint by $f^{min}_{R}(\alpha)$. The ratio between $\Delta f_{L}(\alpha)$ and $\Delta f_{R}(\alpha)$ is denoted by $C$ as follows:
\begin{equation}
\label{eq10}
C=\frac{\Delta f_{L}(\alpha)}{\Delta f_{R}(\alpha)}=\frac{1-f^{min}_{L}(\alpha)}{1-f^{max}_{R}(\alpha)},
\end{equation}
where the index $C$ can be interpreted as a direct measure of the depth of the tail of the spectrum $f(\alpha)$. If $C<1$, the left-hand side is less deep, while if $C>1$, this side is deeper, and if $C=1$, the depths of the tails are the same on both sides. As quoted by \cite{ihlen}, a longer left tail implies that the singularities are stronger, whereas a longer right tail indicates that the singularities are weaker \citep{tanna}.

\subsubsection{The Hurst exponent ($H$)}
According to \cite{defreitas2013}, the exponent $H$ denotes Brownian motion when $H=1/2$, i.e., past and future fluctuations are uncorrelated. However, if $H>1/2$, fluctuations have a tendency to long-term persistence, i.e., an increase in value will most likely be followed by another increase in the short term, and a decrease in value will most likely be followed by another decrease in the short term, while for $H<1/2$, the fluctuations tend not to continue in the same direction but instead turn back on themselves, which results in a less smooth time series \citep{hm}.

\subsection{Sources of multifractality}

Figures \ref{fig1a} to \ref{figsolar} show two other methods, namely, \textit{shuffling} (green dots) and \textit{surrogates} (blue dots), which are used to verify the source of the multifractality. Each multifractality source can be investigated using $h(q)$, which is a function of $q$ (see Eq.~\ref{eq5}), as shown in the right middle panels of Figures \ref{fig1a} to \ref{figsolar}. These behaviours occur in all stars of our sample; thus, the analysis emphasized here is general.

Shuffling a time series destroys the memory signature but preserves the distribution of the data with $h(q)=0.5$ if the source of the multifractality in the time series presents only long-range correlations \citep{defreitas2017}. However, the origin of multifractality can also be due to the presence of fat-tailed probability distributions in the original time series. More specifically, the non-Gaussian (non-linear) effects can be weakened by creating phase-randomized surrogates, thereby preserving the amplitudes of the Fourier transform and the linear properties of the original series by randomizing the Fourier phases \citep{Norouzzadeha,defreitas2017}. In this case, if the source of multifractality is a heavy-tailed distribution that is obtained by the surrogate method, the values of $h(q)$ will be independent of $q$, and $h(q)=0.5$ will not always hold.

\subsection{Effect of fluctuations on the profile of f($\alpha$)}

\begin{figure*}
	\includegraphics[width=0.99\columnwidth]{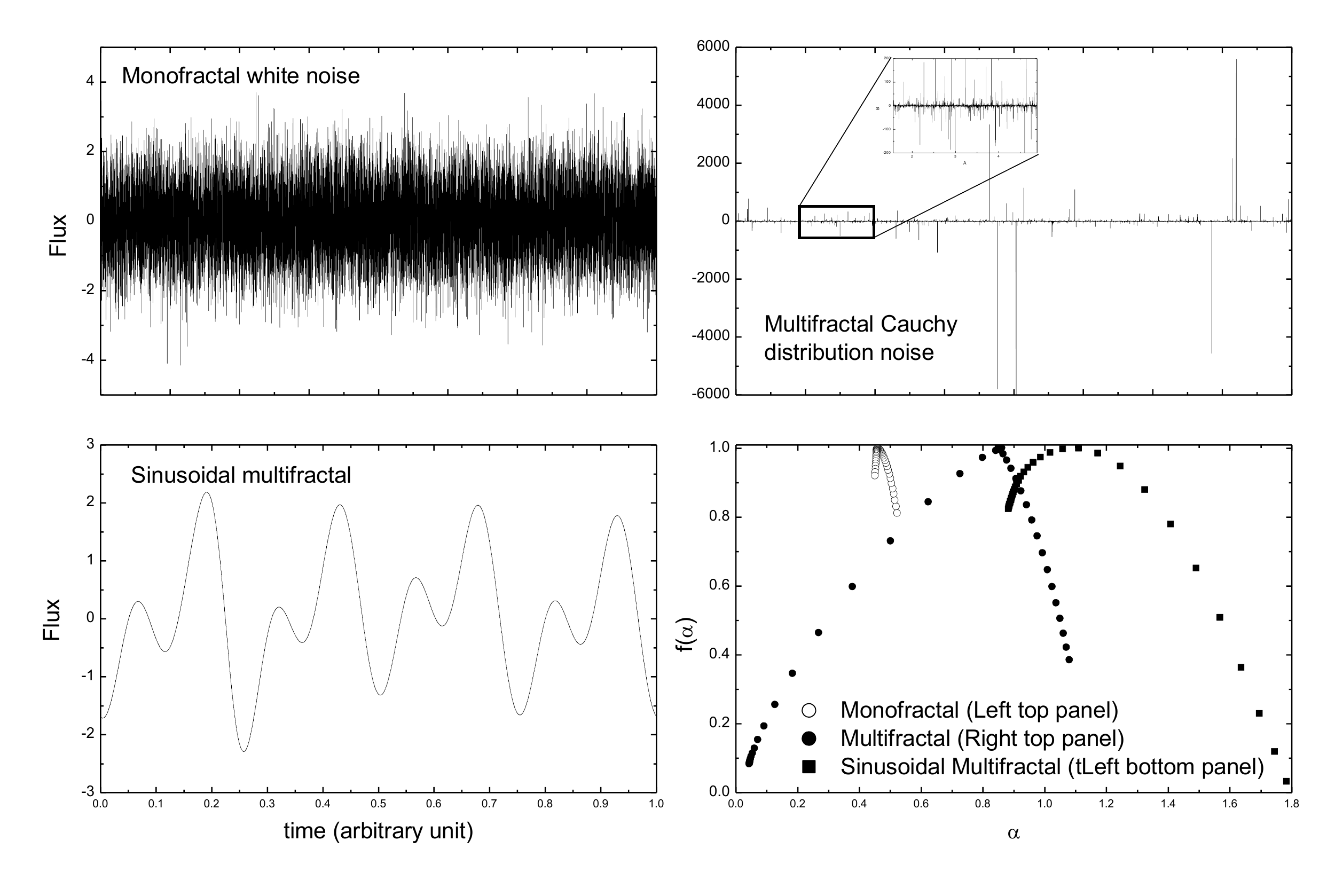}
	\caption{Large and small fluctuations. Left top panel: Monofractal white noise: time-series sample created by random numbers with a normal distribution $(0,1)$. It is expected that the Hurst exponent $H = 0.5$. Right top panel: Random number series with Cauchy distribution. In this case, a spectrum of $H$ is expected. It is clear that a multifractal time series has distinct amplitudes with small and large fluctuations in contrast to monofractal time series. Left bottom panel: Sinusoidal multifractal using two sine functions with frequencies of 4 and 8 Hz. Right bottom panel: Strong left-sided tail of multifractal spectrum $f(\alpha)$ indicates that larger fluctuations are dominant, while a strong right-sided tail indicates that shorter fluctuations are dominant.}
	\label{figflu}
\end{figure*}

Figure \ref{figflu} indicates three types of time series. The first is an uncorrelated-noise time series; therefore, it is expected that the multifractal spectrum is monofractal. In this case, for all scales, $s$, and for all $q$ parameters, we expect a Hurst exponent of $h$=0.5. On the top right panel of the figure, we plot a random number series with a heavy-tailed distribution such as a Cauchy distribution. As shown in this figure, the time series has strong multifractality, likely only because of the distribution. The third time series is a set of two sine functions without noise, containing only semi-sinusoidal variability that is similar to rotational modulation. Its multifractal spectrum is opposite to that found in multifractal Cauchy noise, as mentioned above. According to this figure, the asymmetry in the singularity spectra of time series contains information on the composition of the time series (cf. bottom right panel).

Smaller fluctuations (when $q<0$) correspond to a broad right-sided $f(\alpha)$, i.e., $f(\alpha)$ is more multifractal in its arrangement of smaller fluctuations than in its arrangement of larger fluctuations ($q>0$). The MFDMA method is able to quantify the structure of fluctuations within periods that have small and large fluctuations. Again, Figure \ref{figflu} shows the effect of these fluctuations using multifractal and monofractal time series. In particular, the standard deviation in a monofractal time series does not vary over time. However, in the multifractal time series, the presence of large fluctuations increases the standard deviation in different parts of the time series. In addition, in the presence of noncorrelated background noise, the left tail shrinks; consequently, the signal becomes almost fractal. In contrast, the presence of a strong deterministic signature, e.g., starspots, enlarges the right tail \citep{ihlen,droz}.

In general, different frequencies of small and large fluctuations reveal that the right and left sides of a multifractal spectrum do not belong to the same hierarchical organization. An asymmetrical profile indicates that right- and left-sided multifractal spectra are characterized by an imbalance in fractal complexity between fluctuations of different amplitudes. Indeed, such a profile suggests that variability due to the smaller fluctuations has a hierarchical organization that is different from that of larger fluctuations.

\subsection{Correlation methods: Spearman and Pearson coefficients, and Student $t$-test}
To better understand the level of the relationship among different parameters (e.g., $\Delta\alpha$ and $\Delta\Omega$), we used two bivariate analysis techniques denoted by Pearson’s product moment correlation coefficient $r_{P}$ and Spearman’s rank correlation coefficient $r_{S}$ \citep{press}. Qualitatively, the Pearson method measures the strength of the linear relationship between normally distributed variables. However, the variables cannot be normally distributed or the relationship between the variables is not linear. In this case, the Spearman method is more appropriate than the Pearson method because the former is less sensitive to the presence of outliers and its statistical distribution is independent of the distributions of the two correlated variables \citep{press,mukaka,defreitas2017}. 

Quantitatively, Pearson’s coefficient is given by 
	\begin{equation}
	\label{eq12}
	r_{P}=\frac{S_{xy}}{\sqrt{S_{xx}S_{yy}}},
	\end{equation}
	where $S$ denotes the covariance. The value of $r_{P}$ lies between $-$1 and $+$1, inclusive, where a value of $+$1 denotes a complete positive correlation, $-$1 a complete negative correlation, and a value near zero indicates that the variables $x$ and $y$ are uncorrelated. In cases where the relationship is non-linear, it can sometimes be transformed into a linear one by using the ranks of the variables rather than their actual values. Spearman’s coefficient of rank correlation, denoted by $r_{S}$, can be calculated by applying Eq. (\ref{eq12}) to the ranks, although it is more usual to use the formula given by
	\begin{equation}
	\label{eq11}
	r_{S}=1-\frac{6\sum^{n}_{i=1}d^{2}_{i}}{n(n^{2}-1)},
	\end{equation}
	where $d_{i}$ represents the difference between the ranks of variables $x$ and $y$ (for example, $x=\Delta\alpha$ and $y=\Delta\Omega$) and $n$ is the number of observations. The following $t$-statistic of $r_{S}$ is used to test the null hypothesis
	\begin{equation}
	\label{eq11x}
	t=r_{S}\sqrt{\dfrac{n-2}{1-r^{2}_{S}}},
	\end{equation}
	which is distributed approximately according to Student’s distribution with $n-2$ degrees of freedom \citep{press}.

It is worth mentioning that if the condition $|r_{S}|>|r_{P}|$ occurs, a monotonic relationship is stronger than linear one. Essentially, a monotonic behaviour between the variables can imply a linear relationship, and consequently, the analysis of the results becomes more complex. On the other hand, the presence of outliers may cause this discrepancy between the values of the coefficients. In this case, it is necessary to remove the outliers, recalculate the coefficients and verify any change. To further clarify the correlation, one can use a $t$-statistic to estimate the significance of the correlation coefficient.

For the case of the analysis of the two different distributions, we use the two-sample Student’s $t$-test. For this test, the null hypothesis assumes that the data come from independent random samples from normal distributions with equal means, whereas the alternative hypothesis indicates that the data come from populations with unequal means. In the present study, the $t$-test is calculated using the MATLAB function \texttt{[hyp,p,stats]=ttest2}\footnote{For more details, see https://www.mathworks.com/help/stats/ttest2.html}, where \texttt{hyp} can assume values 0 and 1, \texttt{p} is a probability used to reject or not the null hypothesis, and \texttt{stats} contains information about the test statistic, for instance, the calculated $t$-statistic denoted by $t_{calculated}$ and confidence interval. In particular, if \texttt{hyp=1}, the test rejects the null hypothesis at a given significance level, and if \texttt{hyp=0}, the opposite is true. In the present study, we adopted a significance threshold of 0.001 because significance levels greater than 1$\%$ can lead to wrong conclusions.

We follow the same procedure adopted by \cite{defreitas2017}. First, we define the null hypothesis between the variables $x$ and $y$. Second, a $t$-table is used to find the critical value ($t_{critical}$) based on the degrees of freedom \citep{trauth}. Subsequently, the critical value defines the region on the $t$-test distribution where the null hypothesis can or cannot be rejected. Finally, if and only if the value of the calculated $t$-statistic is greater than $t_{critical}$, the null hypothesis can be rejected in favour of the alternative hypothesis. Otherwise, the null hypothesis is accepted \citep{trauth,press}.

\section{Working sample and data analysis}
The \emph{Kepler} mission performed 17 observational runs for $\sim$90 days each, and these runs, designated as Quarters\footnote{\texttt{http://archive.stsci.edu/pub/kepler/lightcurves/tarfiles/}}, were composed of long-cadence (data sampling every 29.4~min \cite{Jenkins2010}) and short-cadence (sampling every 59~s) observations \citep{van,thompson}; detailed discussions of the public archive can be found in many \emph{Kepler} team publications, e.g., \cite{boru2009,boru}, \cite{batalha}, \cite{koch}, and \cite{basri2011}. Regarding the data format, the \emph{Kepler} archive provides both simple aperture photometry data (SAP) (processed using a standard treatment that removes only the most relevant spacecraft artefacts) and pre-search data conditioning (PDC) data, which are processed using a refined treatment based on the \emph{Kepler} pipeline \citep{Jenkins2010b}. This method removes more thermal and kinematic effects arising from spacecraft operation \citep{van2}.

Based on a working sample adopted by \cite{reinhold} with well-determined rotation periods, we constructed our light curves using only Quarter 3 (Q3) long-cadence data. All time series have been normalized to the median value of Q3. Our final working sample consisted of 662 active stars with physical properties similar to those of the Sun, defined by $T_{\rm eff}$ between 5579 and 5979~K, $\log g$ between 3.94 and 4.94~dex \citep{pen}, and (primary) rotation periods between 24 and 34~days. As proposed by \cite{reinhold}, the active stars are selected using the so-called variability range $R_{var}$. From visual inspection, these authors found that active stars are defined by values of $R_{var}\geq$ 0.003 (3 parts per thousand). In addition, the authors assumed that variability range is a key measure to distinguish between active and inactive stars. The detailed procedure concerning $R_{var}$ can be found in \cite{reinhold}. All of the active stars occupy the dwarf regime, with $\log g>3.5$.

Among the total of 662 stars, 141 were detected to have DR traces, while the rest were identified as having rigid-body rotation. The period interval for the abovementioned selection was based on the results from \cite{lanza}, for which the rotational periods of the Sun are between 24.5~days (equator) and 33.5~days (poles). Values for the rotational periods were estimated using an autocorrelation function and were taken from \cite{reinhold}, and the temperature and gravity were obtained from the \cite{pen} (corrected Sloan Digital Sky Survey [SDSS] temperature and \textit{Kepler} Input Catalogue [KIC] surface gravity). These restrictions were chosen so that our results did not present biases due to mass, temperature or effective gravity. Further details on the final Kepler star sample are shown in Section 3.1.

Our sample also included a dataset of continuous solar observations obtained by Variability of solar Irradiance and Gravity Oscillations (VIRGO) \citep{virgo1,virgo2}. The VIRGO experiment is a component of the payload of the SOHO spacecraft and is based on four instruments, including the Differential Absolute Radiometer (DIARAD), the Luminosity Oscillations Imager (LOI), the PMO6 and a Sun PhotoMeter (SPM). In the present paper, we use the VIRGO data in the green (500 nm) and red (862 nm) bandwidths of the SPM instrument, as proposed by \cite{basri}\footnote{The VIRGO/SPM data can be downloaded from http://www.spaceinn.eu/data-access/calibrated-sohovirgospm-data/}. The VIRGO data analysed in the present work consist of the spectral solar irradiance (SSI) time series with a temporal cadence of 1 min and a date range from April 11, 1996 to March 30, 2014, corresponding to solar cycles 23 and 24 (see Figure \ref{figsolar}). To properly compare these results with the stellar case, we averaged the time series into 30-min cadences to match the \textit{Kepler} measurements. This dataset consists of $\sim$18 years of continuous observations; however, because the temporal window of the \textit{Kepler} Q3 quarter data is $\sim$90 days, we chose a region in the VIRGO/SPM data with few large gaps from April 22, 1999 to July 20, 1999, which was within the Sun’s active phase.

\subsection{Impact of the Kepler pipeline on the multifractal indexes}

To determine the multifractal indexes, we followed the procedure described in the previous section. Table \ref{tab:long} summarizes our results for stars with DR traces. However, we found different behaviour as a function of the Kepler pipeline and, in particular, the procedure adopted by \cite{demedeiros2013} used to remove the outliers.

Our database was extracted from \cite{reinhold}, which uses the PDC pipeline. This pipeline modifies stellar variability on timescales longer than a few days and, therefore, has difficulty distinguishing between astrophysical and instrumental signals. The removal of variability from the light curves can affect the analysis, as mentioned by \cite{gill} (see also Sect. 5.15 of the Kepler Data Characteristic Handbook KSCI-19040-005\footnote{https://archive.stsci.edu/kepler/manuals/Data$\_$Characteristics.pdf}). In general, signals with timescales longer than a few days are expected to be attenuated, affecting the measurement of the multifractal indexes. SAP data are less affected by this attenuation. In this case, we used a 4th-order polynomial fitted to the SAP data to detrend the time series in the Q3 quarter (hereafter called SAP4).

\begin{figure*}
	\includegraphics[width=1.0\columnwidth]{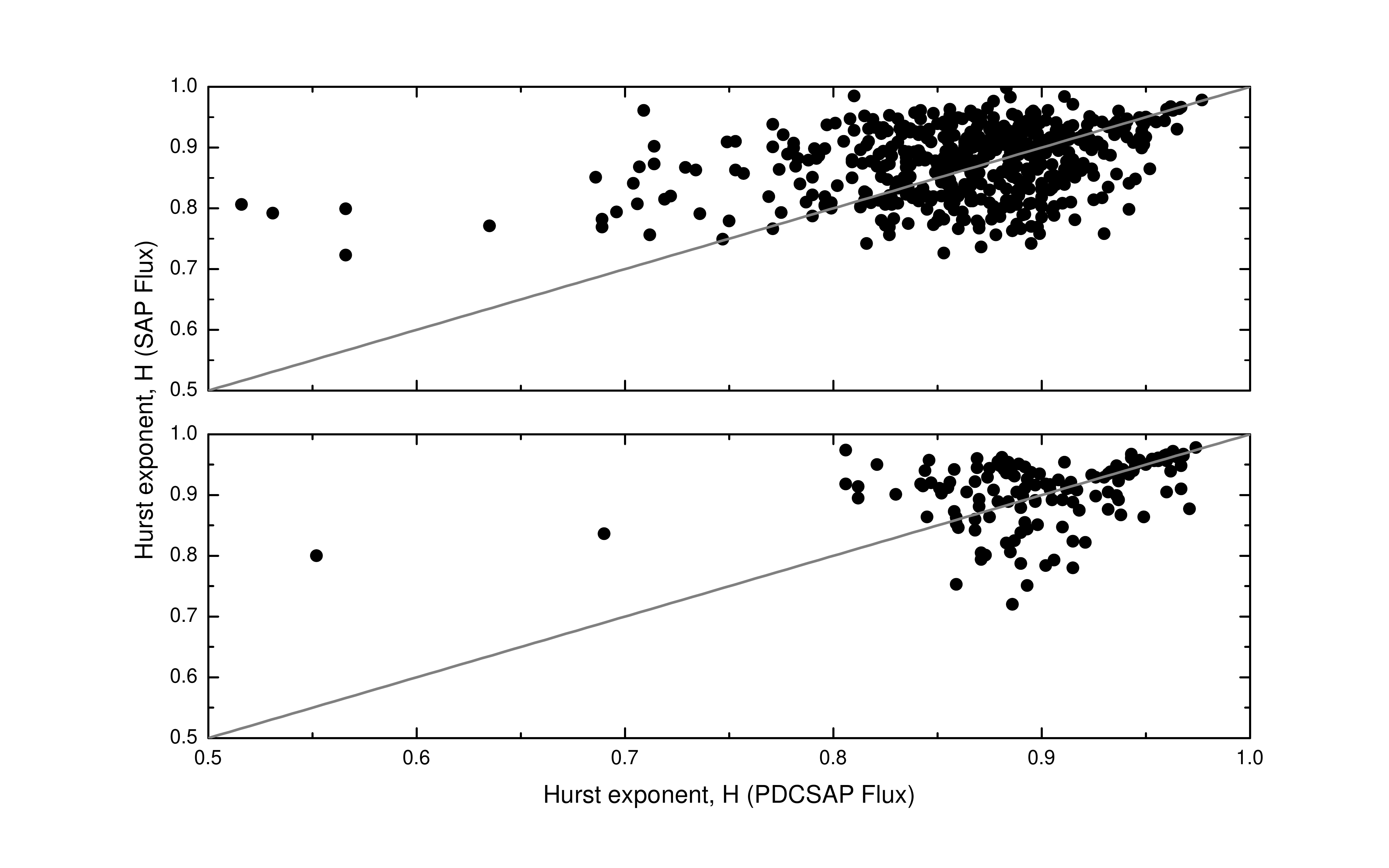}
	\caption{Hurst exponent for PDCSAP flux vs. SAP4 flux in Q3. The top figure shows the correlation for stars without signs of DR, whereas the bottom figure shows the stars with DR traces. The solid line represents the identity between the parameters.}
	\label{figC1}
\end{figure*}

First, we verified the effects of the trends that can be due to stellar variability. For this, we compared the multifractal indexes measured by two types of data: i) PDC data and ii) SAP4 data. Figure \ref{figC1} shows the correlation between these data for index $H$, considering the stars with and without DR traces shown in the bottom and top figures, respectively. In this figure, we found a slight trend in the data; however, in general, the values of $H$ measured by the SAP4 pipeline are higher than those measured by the PDC pipeline. We verified that other multifractal indexes have the same behaviour.

\begin{figure*}
	\includegraphics[width=1.0\columnwidth]{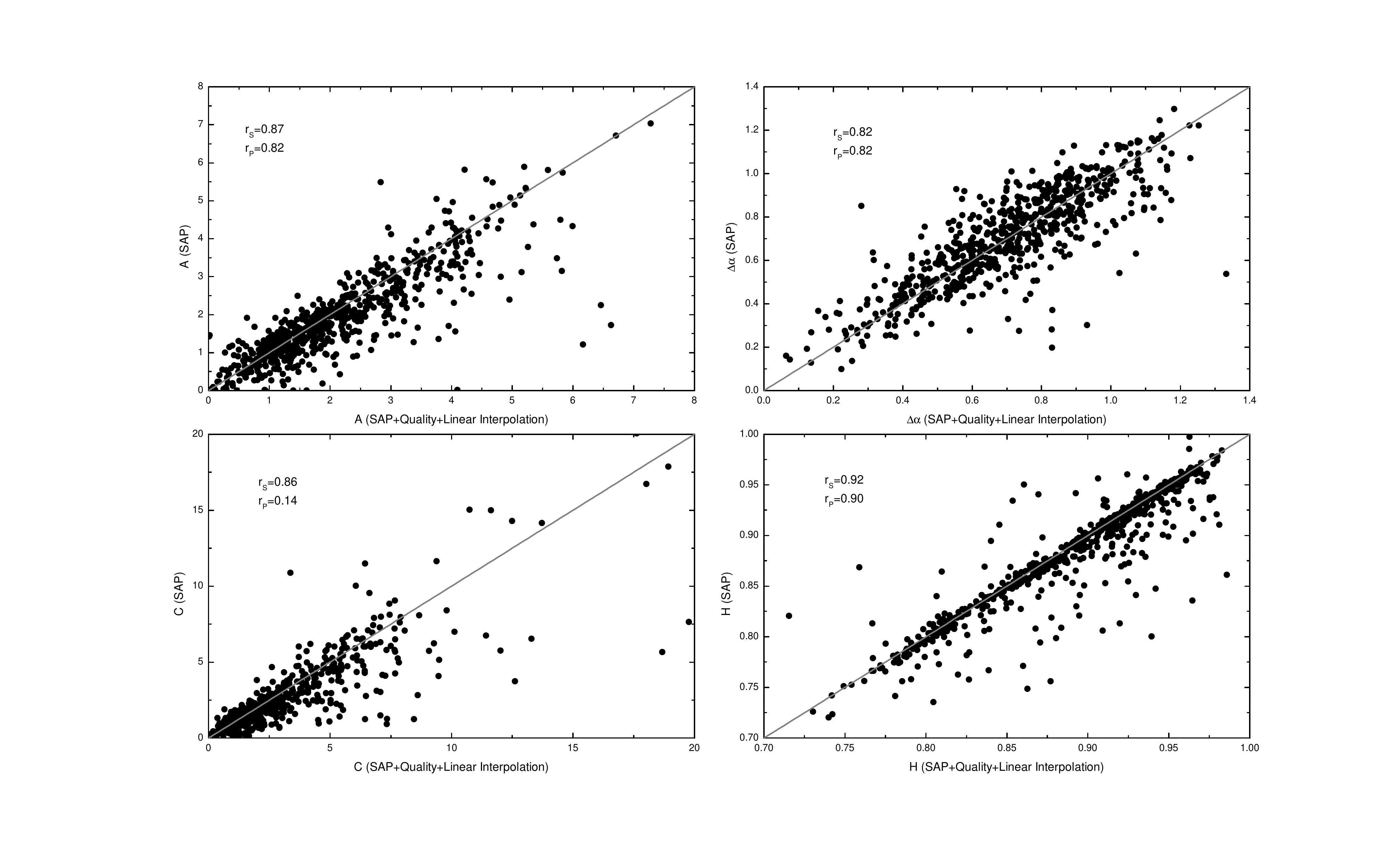}
	\caption{All multifractal indexes for SAP4 flux vs. SAP4QLI flux in Q3. The figure shows the impact of outliers on the multifractal indexes. The solid line represents the identity. The figure also shows the values of the Spearman and Pearson coefficients.}
	\label{figOutlier}
\end{figure*}

The second effect on the \textit{Kepler} pipeline is outliers. This effect should be considered and quantified. To verify this effect, we compared the SAP4 and SAP4+Quality+Linear Interpolation (hereafter SAP4QLI) data. Here, only the data with flag SAP$\_$QUALITY=0 are retained in this analysis. SAP$\_$QUALITY is a flag that contains information on the quality of the data. A linear interpolation is applied to fill in the gaps produced by discarding data with SAP$\_$QUALITY different from zero. Figure \ref{figOutlier} shows that the correlation between the indexes calculated by two types of data is strong, as indicated by the Spearman and Pearson coefficients, $r_{S}$ and $r_{P}$, respectively. In general, we verified that the results obtained in both procedures are in agreement and that the outlier effect can be neglected.

\begin{figure*}
	\includegraphics[width=1.0\columnwidth]{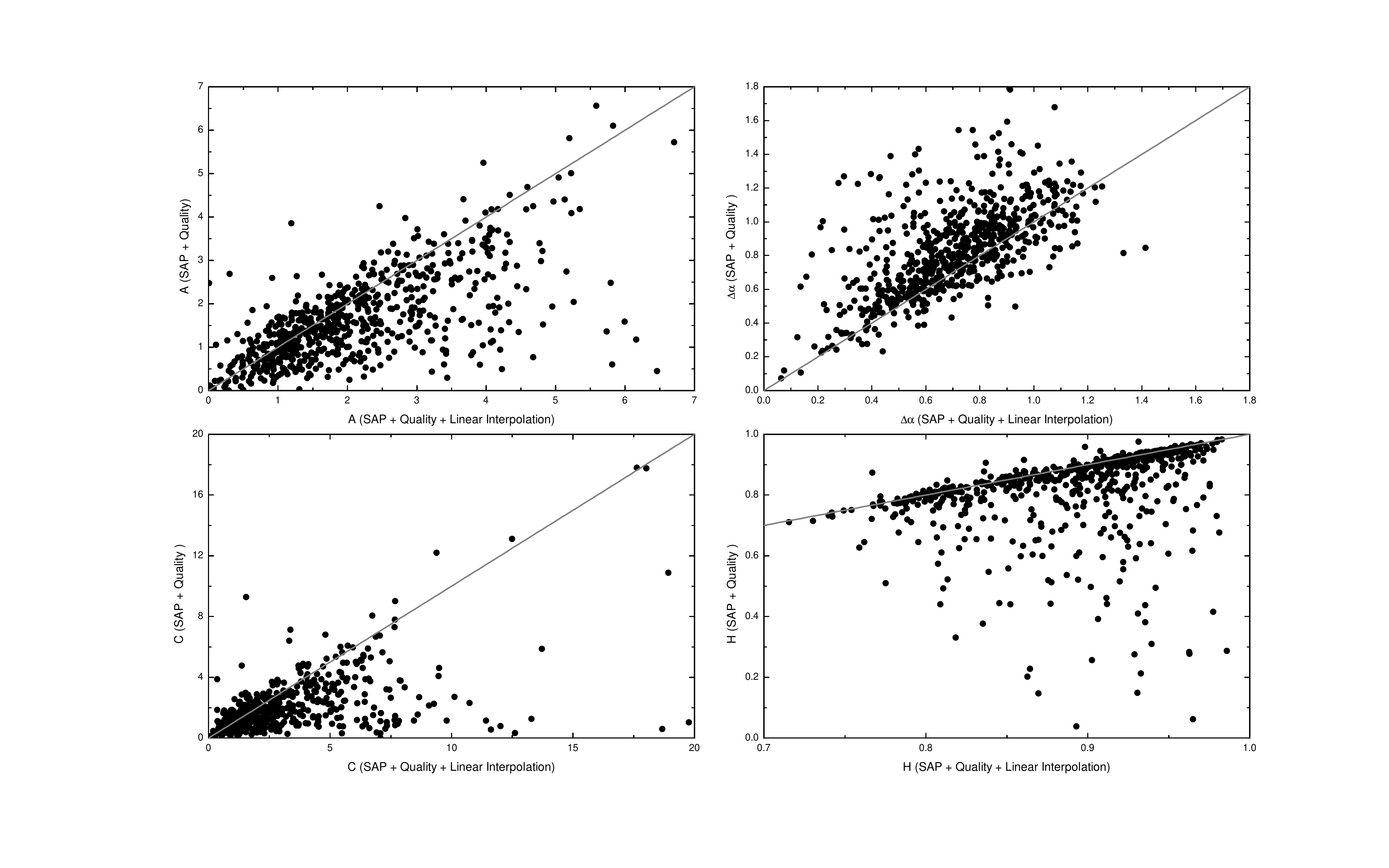}
	\caption{All multifractal indexes for SAP4Q flux vs. SAP4QLI flux in Q3. Figure shows the impact of gaps on the multifractal indexes, considering a linear interpolation. The solid line represents the identity.}
	\label{figGap}
\end{figure*}

Finally, we examined the effect of gaps in the time series after the removal of outliers. We performed a comparison between the SAP4Q and SAP4QLI data. According to the four panels in Figure \ref{figGap}, the four multifractal indexes significantly changed. We verified that the gaps underestimate the values of $A$, $C$ and $H$, whereas the degree of multifractality ($\Delta\alpha$) is overestimated. Generally, the gaps make the time series more complex from a multifractal point of view; therefore, the difference between small and large fluctuations is more pronounced.

\begin{figure*}
	\includegraphics[width=1.0\columnwidth]{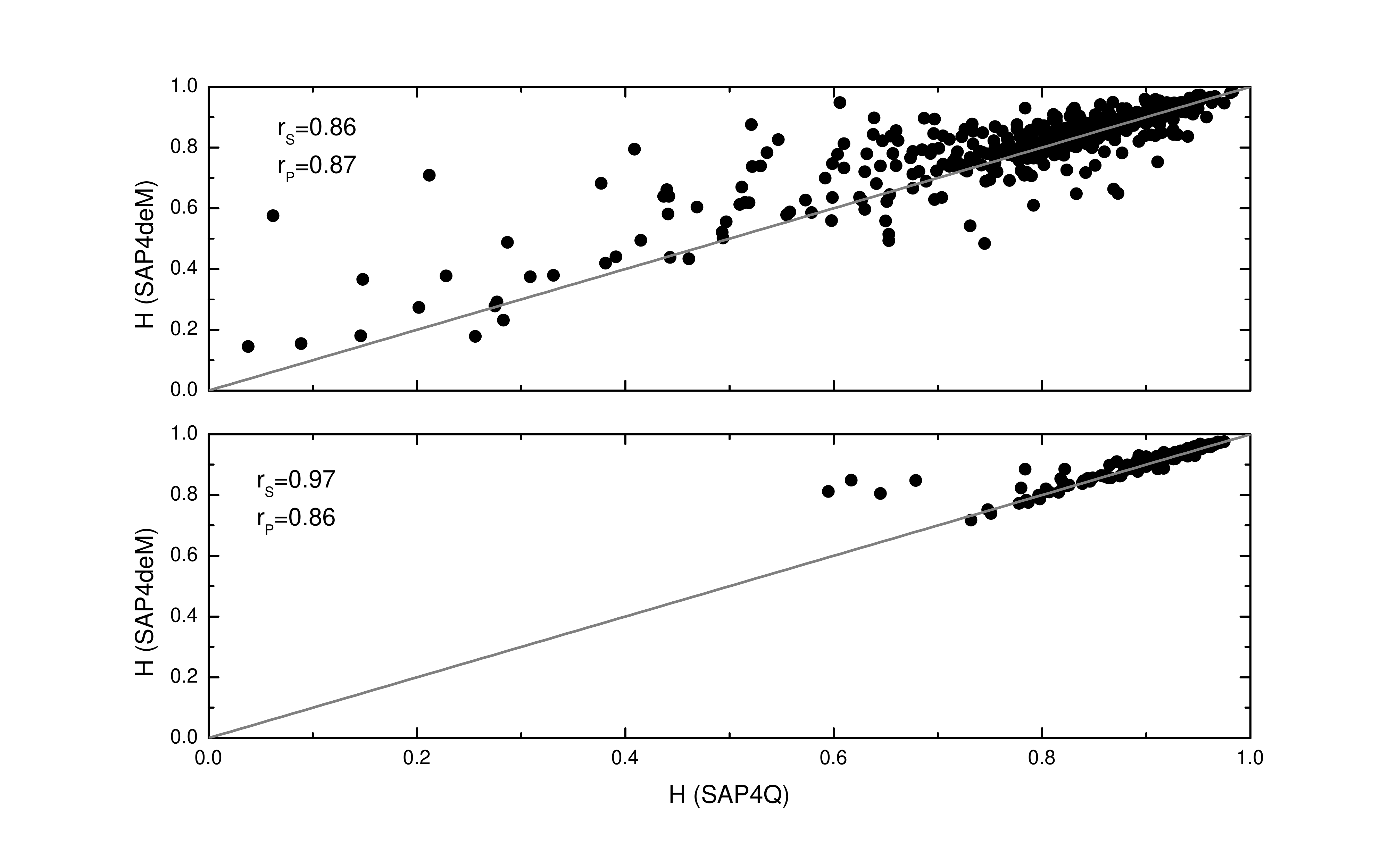}
	\caption{Hurst exponent for SAP4Q flux vs. SAP4deM flux in Q3. The top figure shows the correlation for stars with no DR, whereas the bottom figure shows the correlation for stars with DR traces. The results of the Spearman and Pearson tests are shown at the top of each panel. The solid line represents the identity line.}
	\label{figHsapqDeM}
\end{figure*}

We also examined the effect on the \textit{Kepler} pipeline when another method was used to remove the outliers. The method proposed by \cite{demedeiros2013} was tested for a sample of 4206 CoRoT light curves that present well-defined semi-sinusoidal signatures. Basically, a moving average was used to compute a smoothed time series that was subsequently subtracted to compute the residuals. A 5-$\sigma$ clipping was then applied to remove the outliers as described by \cite{demedeiros2013}. We verified that this method is in good agreement with removal using quality flags. The data obtained by this method were named SAP4deM. Figure \ref{figHsapqDeM} shows a strong correlation between the Hurst exponents of the data obtained by the different methods used to find the outliers.

\begin{figure}
	\includegraphics[width=0.9\columnwidth]{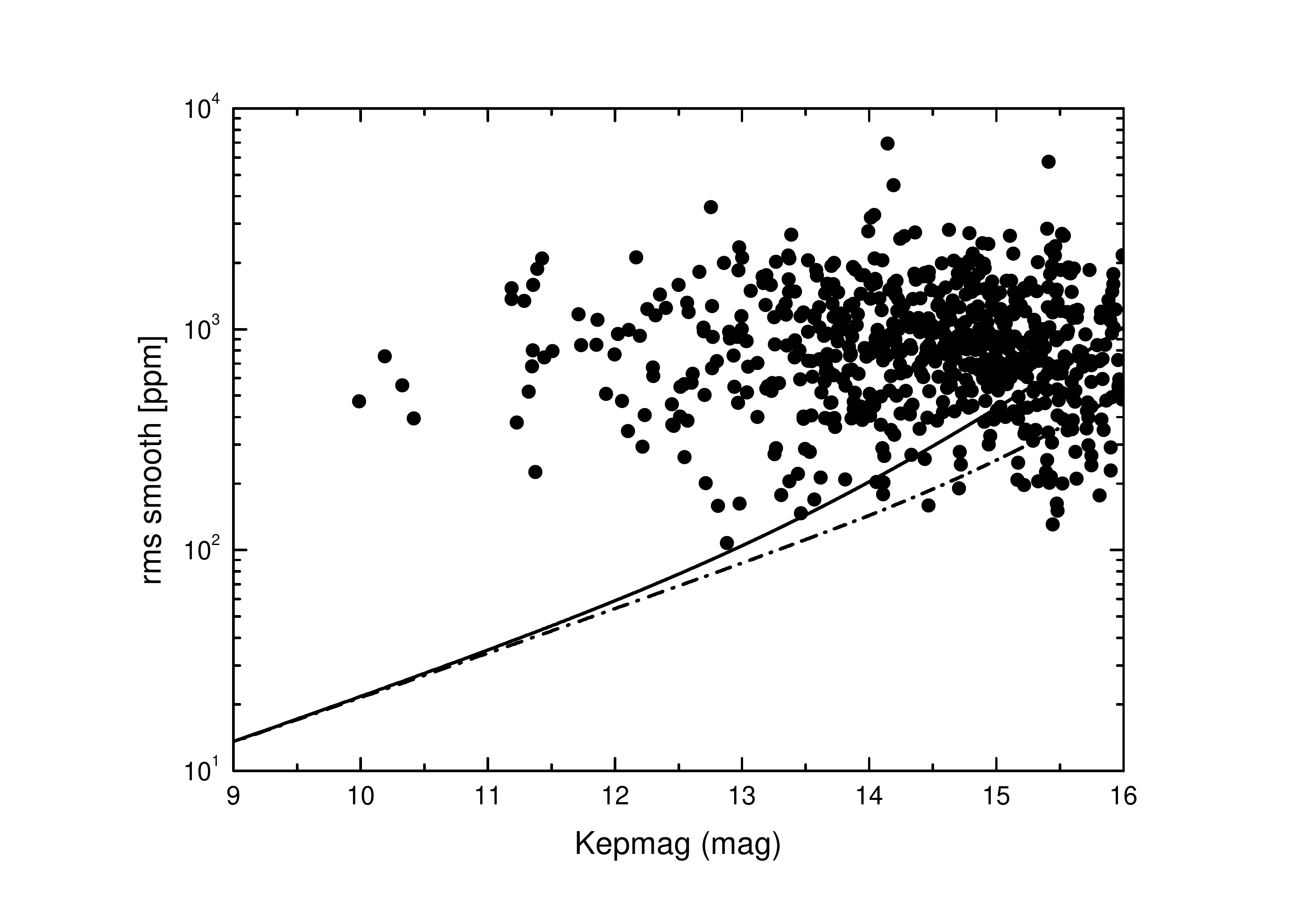}
	\caption{The standard deviation of the smoothed time series for our final sample as a function of the \textit{Kepler} magnitude. The grey dot-dashed and solid lines indicate the lower and upper photon noise levels, respectively.}
	\label{figPNoiseNew}
\end{figure}

\begin{figure*}
	\includegraphics[width=1.0\columnwidth]{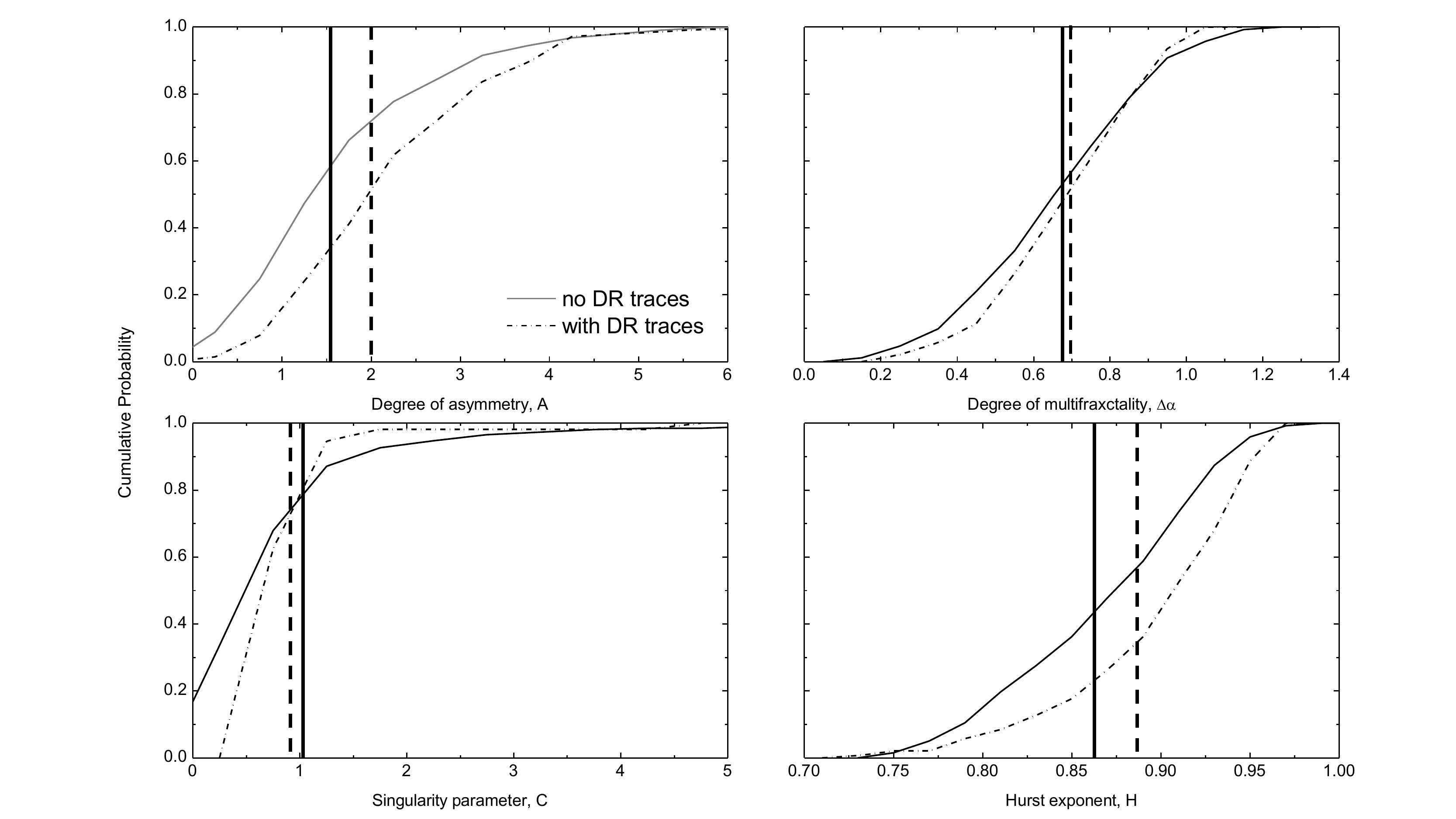}
	\caption{Normalized cumulative probability of four multifractal indexes for stars without (solid line) and with (dot-dashed line) DR traces based on the SAP4 time series. Vertical dashed and solid lines show the medians for stars with and without DR traces, respectively.}
	\label{figCum}
\end{figure*}

\begin{figure*}
	\includegraphics[width=1.0\columnwidth]{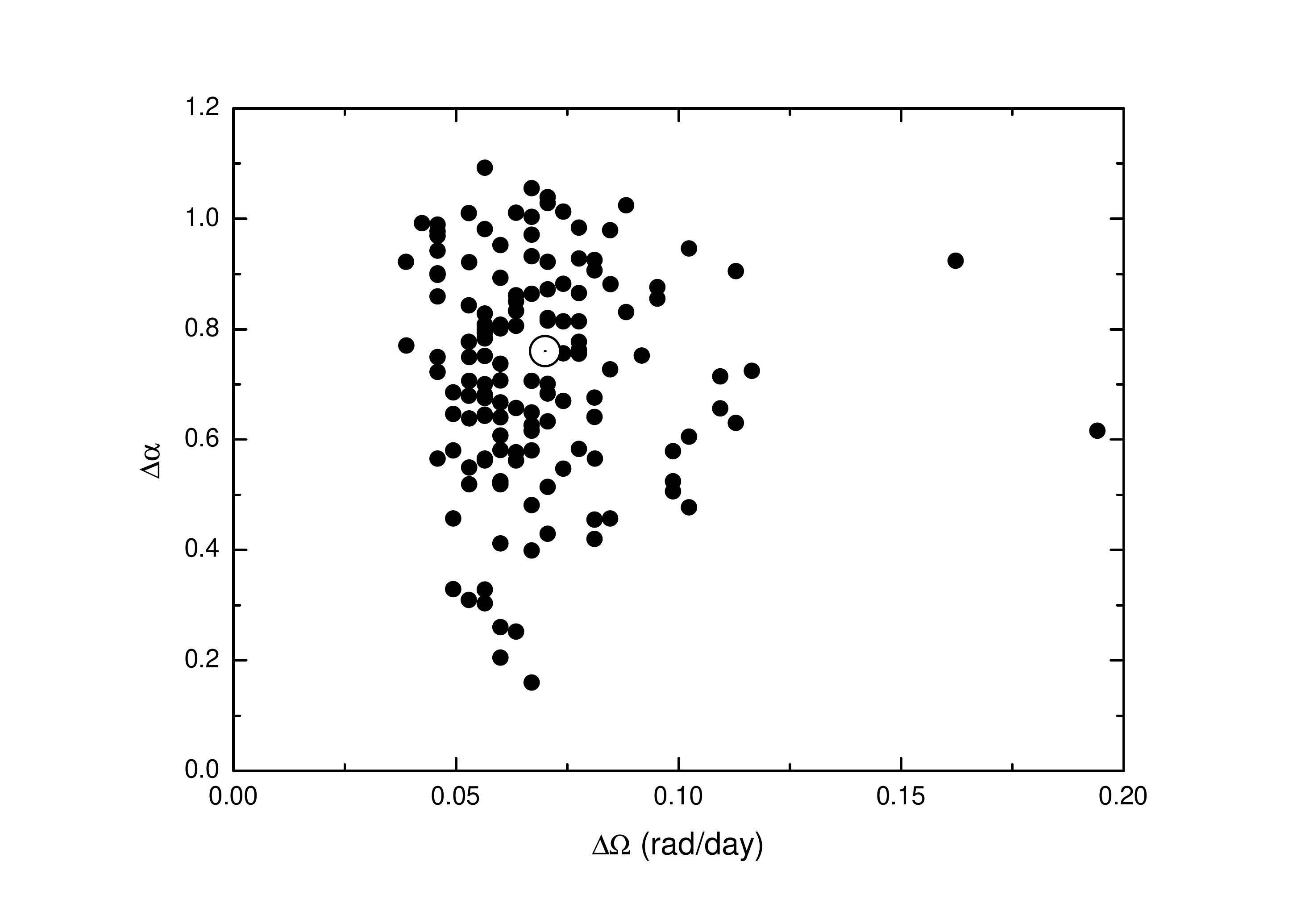}
	\caption{Plot of the degree of multifractality $\Delta\alpha$ versus the absolute horizontal shear $\Delta\Omega$ for Sun-like stars with DR traces identified in the present study. In this figure, SAP4 flux data are used. The Sun is represented by the symbol $\odot$.}
	\label{figDeltaAlfaDeltaOmega}
\end{figure*}

\begin{figure*}
	\includegraphics[width=1.0\columnwidth]{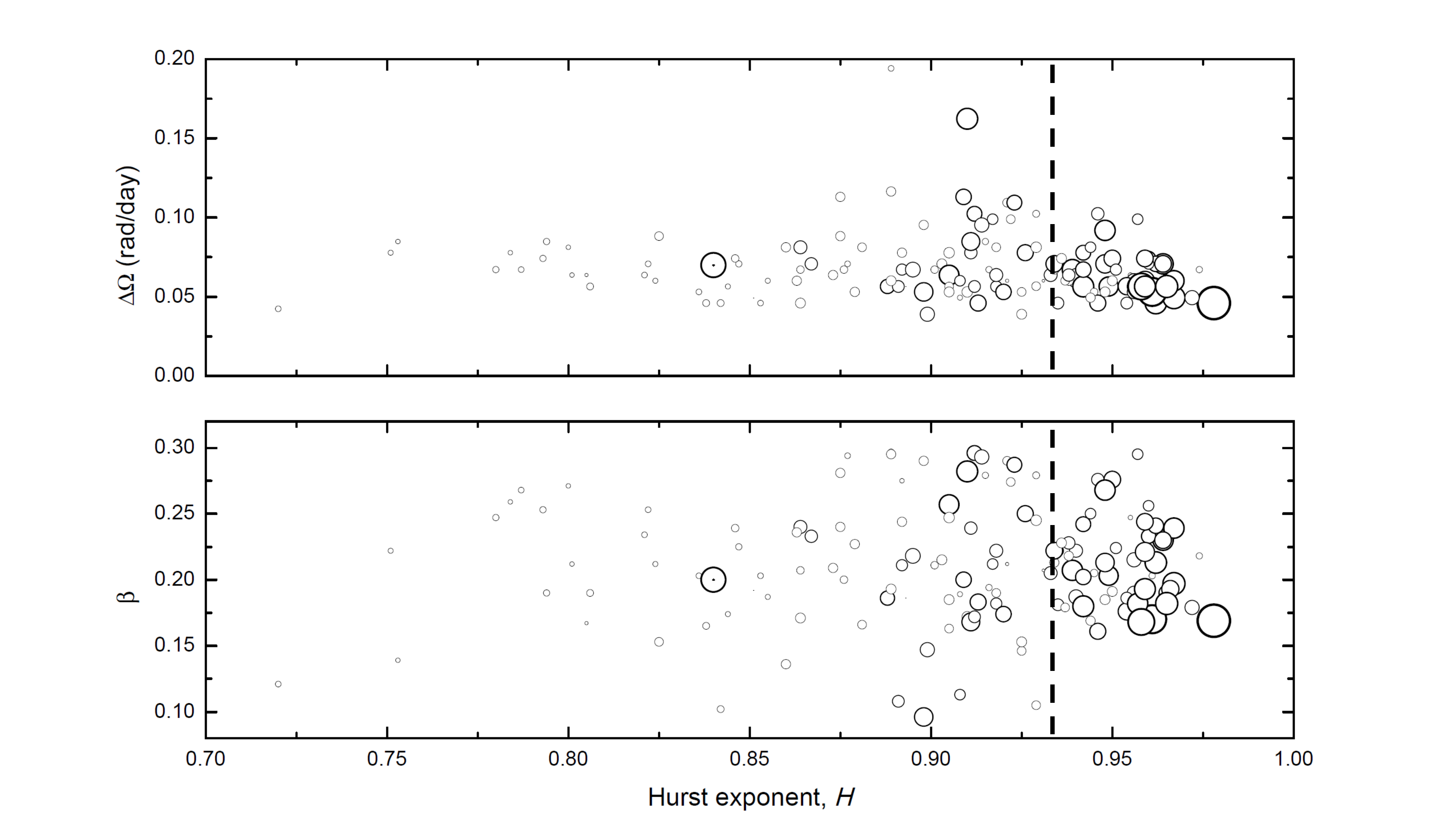}
	\caption{Hurst exponent $H$ vs. relative and absolute shears $\beta$ (bottom) and $\Delta\Omega$ (top), respectively. Both figures show that the two regions of $H$ increase towards more asymmetric multifractal spectra. The dashed line limits these two regimes. Each sample is represented by bubbles of different sizes, which correspond to the intensity of the degree of asymmetry $A$ (see Table~\ref{tab:long}). The dashed line (in $H=0.93$) divides the picture into two domains as a function of $A$, where $\langle A_{H<0.93}\rangle$=1.91 and $\langle A_{H>0.93}\rangle$=3.11. Confidence interval for the difference in sample means is [0.624,1.775].}
	\label{figHvsA}
\end{figure*}

\subsection{Photon shot noise effect}
The data can be sensitive to photon noise. As mentioned by \cite{mathur2014a}, there are several ways to compute the influence of photon noise in data. We have followed the same procedure described by those authors, and we have used the methodology proposed by \cite{Jenkins2010b}. To this end, we calculated the minimum and maximum photon shot noise in the time series of the selected stars. Figure \ref{figPNoiseNew} shows the standard deviation of the smoothed time series as a function of the \textit{Kepler} magnitude using a MATLAB function \texttt{smoothdata}\footnote{For more details, see https://www.mathworks.com/help/matlab/ref/smoothdata.html.} specifically adopted for noisy data. The grey dashed and solid lines indicate the lower and upper photon noise levels, respectively, as defined by \cite{Jenkins2010b}. From Figure \ref{figPNoiseNew}, we conclude that there is no correlation between stellar variability and the estimated magnitude of the stars in the \textit{Kepler} band. Moreover, most of the standard deviations of smoothed time series are above the estimated values for the contribution of photon noise.

\section{Results and discussion}

Mostly, we found that the values of $\alpha_{0}-\alpha_{min}$ are less than $\alpha_{max}-\alpha_{0}$ (see Eq.~\ref{eq8}). Consequently, shorter tails are observed with $q<0$ and $A>1$ (see Table~\ref{tab:long}). This observation means that the fluctuations caused by larger magnitudes are more likely to be monofractals than in the case of $q<0$ (see Figures \ref{fig1a} and \ref{figsolar}, right bottom panels). According to these figures, the abovementioned result is associated with either right or a left truncation, which originates from the levelling of the $q$th-order Hurst exponent for negative or positive values of $q$, respectively \citep{ihlen}.

In the same figures, the left side of $f(\alpha)$ is determined by the positive $q$ values, which filter out larger events, and the opposite applies to its right side. Hence, asymmetry in $f(\alpha)$ reveals non-uniformity, indicating that the corresponding large- and small-scale fluctuations are driven by different mechanisms \citep{droz}. If there is a balance between the mechanisms that control each side of the multifractal spectrum, it will be symmetrical. This equilibrium is broken when one of the fluctuations (small or large) stands out from the other. For example, a sinusoidal signal without background noise produces a wide right tail and a very narrow left tail. A noise-like signal yields the opposite behaviour.

Our results show that the multifractal origins of the rotation behaviour for stars with and without DR are the same; however, there is an important difference. As is the case for stars with no DR traces, the stars with DRs are also related to the long-term persistence measured by the exponent, $H$. However, the right bottom panel of Figure \ref{figCum} shows that the median value of $H$ is lower for stars without DR traces than for stars with such traces. Moreover, the probability of finding high values of $H$ is lower for stars with DR than for stars without DR. This observation suggests that the DR can increase the randomization of data and therefore reduce the values of $H$. This figure highlights other suitable results. Similarly, this process occurs for the distribution of degree of asymmetry, $A$. Previously, for the multifractal indexes, $\Delta\alpha$ and $C$, the distributions had the same profile for both types of stars. For example, what percentage of stars with or without DR have a $A$ that is less than 2? The vertical dashed line at 2 shows us that 70$\%$ of stars without DR and 49$\%$ of stars with DR traces have a $A$ of less than 2, respectively. However, only 30$\%$ of stars without DR have a $A$ of more than 2, compared with 51$\%$ of stars with DR.

In addition, multifractality suggests the existence of different physical mechanisms in the rotational behaviours of stars. Indeed, any active star should show a clear right-handed asymmetry, similar to profile in Figures \ref{fig1a} to \ref{figsolar}. The opposite behaviour is expected for inactive stars \citep{defreitas2019}.

We also analysed the behaviour of the degree of multifractality, $\Delta\alpha$, as a function of the relative absolute shear, $\Delta\Omega$, for our sample of 141 stars despite the narrow range of rotational periods considered in this study, i.e., from 24.5 to 33.5 days. Figure \ref{figDeltaAlfaDeltaOmega} shows the behaviour of $\Delta\alpha$ versus $\Delta\Omega$, from which we can observe a soft correlation of increasing $\Delta\alpha$ towards stronger DR traces. For instance, as shown by \cite{daschagas2016}, the relative DR shear ($\Delta P/P$) increases with longer rotation periods, which is in agreement with previous observations \citep{reinhold} and theoretical approaches \citep{kuker}. In conclusion, we found a slight trend between $\Delta\alpha$ and $\Delta\Omega$, and stars with stronger DR ($\Delta\alpha>0.6$) seem to be more complex than stars with weaker DR.

We estimate that the correlation between $\Delta\alpha$ and $\Delta\Omega$ can be neglected, since the Pearson and Spearman correlation coefficients are very weak, $r_{P}=0.007$ and $r_{S}=0.003$, respectively. In other words, this result does not allow us to reject the null hypothesis that there is no correlation between $\Delta\alpha$ and $\Delta\Omega$. The two-sided $t$-statistic also reveals that the null hypothesis cannot be rejected in favour of the alternative hypothesis, since as $-2.576<t_{calculated}(0.036)<2.576$ at the $1\%$ significance level. Here, 2.576 is the critical value $t_{critical}$.

In Figure \ref{figHvsA}, we plot $\beta$ and $\Delta\Omega$ as a function of the Hurst exponent, $H$, for each of the 141 Sun-like stars with the DR traces that were identified in \cite{reinhold}. Here, $\beta:=|P1-P2|/\max\{P1,P2\}$, and the absolute horizontal shear is defined as $\Delta\Omega:=2\pi/|1/P1-1/P2|$.

The overall distributions of $\beta$ and $\Delta\Omega$ in Fig. \ref{figHvsA} show two regimes separated by a dashed line. The stars in the two regimes are shown with different sizes to indicate that more asymmetric multifractal spectra are found to the right of the dashed line and less asymmetric ones are localized to the left of the line. In this figure (top panel), the distributions of $H$ for values greater than 0.93 indicate that the broadness of $\Delta\Omega$ is narrower. This result means that the difference between periods $P1$ and $P2$ is small. In contrast, for the same regime of $H$, $\beta$ is spread more widely; regardless, the highest values of $A$ are found there. In addition, the regime of $H<0.93$ shows a maximum scatter of about a factor of 3 for parameters $\beta$ and $\Delta\Omega$. In addition, this result shows that for a star rotating closer to a rigid body, the asymmetry between the tails of the multifractal spectrum is stronger; therefore, weak singularities ($q<0$) are dominant in the signal, indicating that the star has a weak magnetic activity on its surface.

Our main hypothesis is to determine whether the degree of asymmetry, $A$, is a segregation factor. Thus, as the restriction of the period to the solar values (as described in Section 3) weakens the correlation between the rotation period and the Hurst exponent, the impact of the degree of asymmetry on the correlations between $H$ and $\Delta\Omega$ and $\beta$ can be attenuated.

To test this hypothesis, we separate the parameters $A$ into two populations with known standard deviations. The first sample contains the values of $A$ for $H<0.93$, while the second is defined for the values of $H>0.93$. The value found of \texttt{hyp}=1 indicates that the null hypothesis can be rejected at the 0.1$\%$ significance level. As a result, we found that the distributions of $A$ are from populations with unequal means, and therefore the two samples do not come from a common distribution. To reinforce this result, a one-sided test was done. The $t$-statistic reveals that, as $ t_{calculated}(7.003)> t_{critical}(3.291)$, the null hypothesis can be rejected in favour of the alternative hypothesis, which assumes that the means are different. In this case, we conclude that the degree of asymmetry $A$ can be considered a segregation factor for both correlations shown in Figure \ref{figHvsA}.

\subsection{Origin of multifractality in the working sample}

As cited in Section 2.2, multifractality can have two sources: long-range correlations (memory) and a fat-tailed probability distribution. In the case of astrophysical time series, \cite{defreitas2016,defreitas2017} showed that the first source of multifractality applies and that the second source has no effect because the probability distribution is roughly Gaussian.

In the vast majority of our sample stars, the multifractal spectrum showed not only that the time series of active \textit{Kepler} stars and the Sun are multifractals but also that they have strong right-handed asymmetry, i.e., the degree of asymmetry, $A$, exceeds unity (see Figures \ref{fig1a} to \ref{figsolar}). The right side of $f(\alpha)$ is determined by the negative $q$ values, which filter out smaller events, and the opposite applies to its left side. As mentioned by \cite{droz}, this behaviour indicates that multifractality operates on small-scale fluctuations (seen through $q<0$), whereas the dynamics of large-scale fluctuations ($q>0$) are reduced in this context. Droz et al. still emphasize that asymmetries found in the multifractal spectrum can provide important insights into the series structure.

We verified that the randomization destroyed the correlations, causing the results to closely resemble those for white noise (see first panel in Figure \ref{figflu}). In addition, the multifractal spectra (see Figures \ref{fig1a} to \ref{figsolar}) reveal that the green curves are closer to $h=0.5$. Moreover, our time series do not present significant non-Gaussian effects. Furthermore, the surrogate time series is very similar to the original series, as shown in Figures \ref{fig1a} and \ref{fig1b} (bottom right, blue curves). Thus, we conclude that the multifractality of our sample is actually due to correlations that physically originated from the dynamics of the starspots.

\subsection{Origin of the tails of the multifractal spectra}

For the sake of clarity, Figures \ref{figFsq} shows only the fluctuation functions, $F_{q}(s)$ as a function of scale with size $s$ for the original KIC002450531 time series. We also show the different types of data as mentioned by Figures \ref{fig1a} to \ref{fig1a4}. In this way, the reader can understand the behaviour of the fluctuation functions in further detail. For negative $q$s, the spread of $F_{q}(s)$ is considerably increased. Consequently, the width of the right tail of the multifractal spectrum is increased. Undoubtedly, a relevant question is the origin of the tail in the multifractal spectrum.

According to \cite{droz}, an alternative way to illustrate the origin of the tail of the multifractal spectrum, $f(\alpha)$, is to use a binomial cascade model with a Gaussian $N(\lambda,\sigma)$ distribution defined by the mean, $\lambda$, and variance, $\sigma$. One binomial cascade already accounts for the symmetric spectrum, just as when the spectrum is a result of the average of the spectra produced by each cascade, considering different $\lambda$ and fixed $\sigma$ \citep[see Figure 2 from][]{droz}.

In the model presented by the authors, they concluded that the left-sided asymmetry is observed because the small fluctuations are dominant in such cascades and therefore approach the central limit theorem (white noise), which reduces the right tail to a monofractal. This process destroys the complexity on the level of small fluctuations more so than on the level of the large fluctuations \citep{ihlen,droz}. The theorem considers that when independent random variables are added (in the present case, superimposing cascades with abundances of small fluctuations), their accurately normalized sum yields a Gaussian distribution. Indeed, this effect occurs even if the distributions of the original variables are non-Gaussian \citep{kwag}.

In summary, the smaller fluctuations produce a strong multifractal hierarchical structure, whereas the larger fluctuations are reduced towards a monofractal structure. In the case of right-sided asymmetry, the opposite process occurs, i.e., a larger level of complexity for the smaller fluctuations and more noise-like features for the larger fluctuations. Nevertheless, this procedure is not easy and requires refinement as described below.

\cite{droz} developed a technique that adds the randomization of the larger fluctuations to the binomial cascade model for $\lambda=1.05$. Those investigators selected events larger than $4.7\sigma$ (approximately 1$\%$ of the sample) of the $\lambda=1.05$ model. Afterwards, their values were replaced by the sum of events larger than $4.7\sigma$ and for $N(0,1)$ white noise. The final result of this construction preserves the structure of the original small fluctuations and randomizes the largest ones.

We simulated multifractal data using the code developed by \cite{wen}. Figure \ref{figflu} is the result of this compilation. This code\footnote{The code is available at \texttt{http://www.mathworks.com/
matlabcentral/fileexchange/29686-multifractal-model-of-asset-returns-mmar}} allows the creation of a time series based on the multiplicative lognormal cascade, similar to the process used by \cite{droz}. We adopted the same parameters and procedures cited by \cite{droz} in which $\lambda$ equals 1.05 and 1.5 to analyse the left-sided asymmetry, whereas for the right-sided asymmetry, only $\lambda$=1.05 was used. $\sigma$=1 was used for both cases. The final time series comprised two parts: i) the original cascade series and ii) white noise. As shown in Figure 4 from \cite{droz}, large fluctuations are more abundant than small fluctuations, and therefore, the degree of asymmetry is positive, confirming that the hierarchical structure was reversed as intended. Thus, the model applied here corroborates the conclusions found by \cite{droz}.

From the multifractal perspective, the rotational signature due to the lifetime of active regions has a dual nature. The results above indicate that the magnetic fields of stars in the active phase seem to function via two mechanisms operating on different scales. In addition, stars in the inactive phase have less complex magnetic fields than active ones. However, the inactive stars exhibit a reduced or absent starspot signature and strong background noise. A more detailed study would be necessary to verify whether this difference is real or merely a statistical artefact.

\section{Final remarks}
First, we conclude that an analysis that incorporates the multifractal method can add diagnostic power to contemporary methods of time-series analysis for studying the signatures of rigid bodies and those with DR. We suggest an alternative explanation for the effects of DR in Sun-like stars that is based on its multifractal nature. We also suggest that the DR measured by parameters $\beta$ and $\Delta\Omega$ for the stars that were studied here is linked to the degree of asymmetry. This result was possible because we restricted the interval of rotation period to one solar period and thus eliminated the strong correlation between the exponent, $H$, and period \citep{defreitas2013, defreitas2016, defreitas2017}. Consequently, the degree of asymmetry, $A$, is the second multifractal index that is related to the effect of the rotational modulation on the photometric time series. We also verified that the degree of multifractality, $\Delta\alpha$, is associated with the DR. Figure \ref{figDeltaAlfaDeltaOmega} shows the behaviour of $\Delta\Omega$ versus $\Delta\alpha$, from which we can observe a soft trend of increasing $\Delta\alpha$ as DR becomes stronger. In general, it indicates that our sample is not affected by the level of complexity of the time series, as measured by $\Delta\alpha$.

The first impactful result of this work is the suggestion that starspots for time series with and without DR have distinct dynamics. In addition, our approach suggests that the DR signature and rigid-body rotation period are explicitly governed by local fluctuations with smaller magnitudes. Additionally, we identified the overall trend whereby the DR, which is represented by parameters $\beta$ and $\Delta\Omega$, is distributed in two $H$ regimes segregated by the degree of asymmetry, $A$.

The second impactful result concerns the fact that the magnetic fields of active stars are apparently governed by two mechanisms with different levels of complexity for fluctuations. Similar to results obtained by \cite{droz} to the singularity spectra with a right asymmetry, the present study revealed that the behaviour for generating right-sided asymmetry is the inverse of that for left-sided asymmetry. This inversion is characterized by a higher level of complexity for smaller fluctuations and a more noise-like profile for larger fluctuations.

The results of the present paper are not the final results of the analysis of stellar rotation as a multifractal process. Indeed, there is an outstanding question regarding the multifractal behaviours present in the \textit{Kepler} time series that can motivate further research. We are examining a large sample of stars in the inactive phase. It is theoretically expected that stars in this stage exhibit inverse behaviour to that found in active stars, i.e., the asymmetry of the multifractal spectrum is left. One hypothesis to be tested is whether the exponent $H$ for inactive stars is correlated to the value of $H$ measured when these stars are in the active phase. If so, we will have a method for calibrating the Hurst exponent and thus for inferring the rotation periods of inactive stars.

In summary, the observed multifractality is primarily a consequence of such correlated inputs that generate a heterogeneous multifractal output. In this case, the multifractality is a result of magnetic activity control mechanisms for activity-related long-term persistent signatures. From an astrophysical perspective, the detection of multifractal signatures in the \textit{Kepler} time series is of interest because it indicates that the control mechanisms regulating the magnetic activity may interact on different timescales in a system operating far from equilibrium. Furthermore, these results indicate that stellar magnetic activity behaviour is even more complex than models can extract using the Fourier transform or Gaussian processes. Finally, this paper poses a new theoretical effort to develop statistical models of stellar rotation and the processes within starspot formation.

\section*{Acknowledgements}

DBdeF acknowledges financial support 
from the Brazilian agency CNPq-PQ2 (grant No. 311578/2018-7). Research activities of STELLAR TEAM of Federal University of Cear\'a are supported by continuous grants from the Brazilian agency CNPq. JRM acknowledges CNPq, CAPES and FAPERN agencies for financial support. This paper includes data collected by the \textit{Kepler} mission. Funding for the \textit{Kepler} mission is provided by the NASA Science Mission directorate. All data presented in this paper were obtained from the Mikulski Archive for Space Telescopes
(MAST).





\begin{thebibliography}{99}
\bibitem[\protect\citeauthoryear{Affer et al.}{2012}]{affer}Affer L., Micela G., Favata F., Flaccomio E., 2012, MNRAS, 424, 11

\bibitem[\protect\citeauthoryear{Aigrain et al.}{2015}]{aigrain}
Aigrain S. et al., 2015, MNRAS, 450, 3211


\bibitem[\protect\citeauthoryear{Aschwaden}{2011}]{a2011}
Aschwanden, M. J. 2011, Self-Organized Criticality in Astrophysics. The Statistics
of Nonlinear Processes in the Universe, Springer-Praxis: New York

\bibitem[\protect\citeauthoryear{Basri et al.}{2011}]{basri2011}
Basri et al., 2011, \aj, 141, 20

\bibitem[\protect\citeauthoryear{Basri, Walkowicz \& Reiners}{2013}]{basri}Basri, G., Walkowicz, L. M., \& Reiners, A. 2013, ApJ, 769, 37

\bibitem[\protect\citeauthoryear{Batalha et al.}{2010}]{batalha}
Batalha, N. M., Borucki, W. J., Koch, D. G., et al. 2010, ApJL,
713, L109

\bibitem[\protect\citeauthoryear{Borucki et al.}{2009}]{boru2009}Borucki, W., Koch, D., Batalha, N., et al. 2009, IAU Symposium, 253, 289

\bibitem[\protect\citeauthoryear{Borucki et al.}{2010}]{boru}Borucki W. J. et al., 2010, Science, 327, 977

\bibitem[\protect\citeauthoryear{Cordeiro \& de Freitas}{2019}]{defreitas2019}Cordeiro, J. G., \& de Freitas, D. B., 2019, in preparation

\bibitem[\protect\citeauthoryear{Das Chagas et al.}{2016}]{daschagas2016}
Das Chagas, M. L., Bravo, J. P., Costa, A. D., Ferreira Lopes, C. E., Silva Sobrinho, R., Paz--Chinchon, F. et al. 2016, MNRAS, 463, 1624

\bibitem[\protect\citeauthoryear{De Medeiros et al.}{2013}]{demedeiros2013}
De Medeiros, J. R., Lopes, C. E. F., Le\~ao, I. C., et al. 2013, \aap, 555, 63

\bibitem[\protect\citeauthoryear{de Freitas \& De Medeiros}{2009}]{defreitas2009}
de Freitas, D. B., \& De Medeiros, J. R. 2009, Europhys. Lett, 88, 19001

\bibitem[\protect\citeauthoryear{de Freitas et al.}{2013a}]{defreitas2013b}
de Freitas, D. B., Fran\c{c}a, G. S., Scherrer, T. M., Vilar, C. S., \& Silva, R. 2013a, Europhys. Lett., 102, 39001

\bibitem[\protect\citeauthoryear{de Freitas et al.}{2013b}]{defreitas2013}
de Freitas, D. B., Le\~ao, I. C., Lopes, C. E. F., De Medeiros, J. R., et al. 2013b, ApJL, 773, L18

\bibitem[\protect\citeauthoryear{de Freitas et al.}{2016}]{defreitas2016}
de Freitas, D. B., Nepomuceno, M. M. F., de Moraes Junior, P. R. V., Lopes, C. E. F., Le\~ao, I. C. et al. 2016, ApJ, 831, 87

\bibitem[\protect\citeauthoryear{de Freitas et al.}{2017}]{defreitas2017}
de Freitas, D. B., Nepomuceno, M. M. F., de Moraes Junior, P. R. V., Lopes, C. E. F., Le\~ao, I. C. et al. 2017, ApJ, 843, 103

\bibitem[\protect\citeauthoryear{Drozdz \& Oswiecimka}{2015}]{droz}Drozdz, S. \& Oswiecimka, P., 2015, Physical Review E, 91, 030902

\bibitem[\protect\citeauthoryear{Donner \& Barbosa}{2008}]{db2008}
Donner, R. V., \& Barbosa, S. M. 2008, Nonlinear Time Series Analysis in the Geosciences--Applications in
Climatology, Geodynamics and Solar-Terrestrial Physics (Berlin: Springer)

\bibitem[\protect\citeauthoryear{Elia et al.}{2018}]{elia} Elia, D., Strafella, F., Dib, S., Schneider, N. et al., 2018, MNRAS, 481, 509

\bibitem[\protect\citeauthoryear{Feder}{1988}]{feder1988}
Feder, J. 1988, Fractals, Plenum Press, New York    


\bibitem[\protect\citeauthoryear{Fr{\"o}hlich et al.}{1995}]{virgo1}
Fr{\"o}hlich, C., Romero, J., Roth, H., Wehrli, C., Andersen, B. N., et al., 1995, Sol. Phys., 162, 101

\bibitem[\protect\citeauthoryear{Fr{\"o}hlich et al.}{1997}]{virgo2}
Fr{\"o}hlich, C., Crommelynck, D., Wehrli, C., Anklin, M., Dewitte, S. et al., H.~J. 1997, Sol. Phys., 175, 267

\bibitem[\protect\citeauthoryear{Gilliland et al.}{2015}]{gill} Gilliland, R. L., Chaplin, W. J., Jenkins, J. M., Ramsey, L. W., \& Smith, J. C. 2015, AJ, 150, 133


\bibitem[\protect\citeauthoryear{Gu \& Zhou}{2010}]{gu2010}
Gu, G.-F., \& Zhou, W.-X. 2010, Phys. Rev. E, 82, 011136


\bibitem[\protect\citeauthoryear{Hampson \& Mallen}{2011}]{hm}
Hampson, K. M., \& Mallen, E. A. H. 2011, Biomedical Optics Express, 2, 464


\bibitem[\protect\citeauthoryear{Hurst}{1951}]{hurst1951}
Hurst, H. E. 1951, Trans. Am. Soc. Civ. Eng., 116, 770

\bibitem[\protect\citeauthoryear{Hurst, Black \& Simaika}{1965}]{hurst1965}
Hurst, H. E. \& Black, R. P., \& Simaika, Y. M. 1965, Long-term storage: an experimental study, Constable, London

\bibitem[\protect\citeauthoryear{Ihlen}{2012}]{ihlen}
Ihlen, E. A. F. 2012, Front. Physiology 3, 141

\bibitem[\protect\citeauthoryear{Ivanov et al.}{1999}]{ivanov1999}
Ivanov, P. Ch., Amaral, L. A. N., Goldberger, A. L., et al. 1999, \nat, 399, 461

\bibitem[\protect\citeauthoryear{Jenkins et al.}{2010a}]{Jenkins2010}
Jenkins et al., 2010a, \apjl, 713, L87

\bibitem[\protect\citeauthoryear{Jenkins et al.}{2010b}]{Jenkins2010b}Jenkins, J. M., Caldwell, D. A., Chandrasekaran, H., et al. 2010b, ApJL, 713, L120


\bibitem[\protect\citeauthoryear{Kantelhardt et al.}{1999}]{Kantelhardt}
Kantelhardt, J.W., Zschiegner, S.A., Koscienlny-Bunde, E., \& Havlin, S. 2002, Physica A, 316, 87

\bibitem[\protect\citeauthoryear{Koch et al.}{2010}]{koch}Koch, D. G., Borucki, W. J., Basri, G., et al. 2010, ApJL, 713, L79

\bibitem[\protect\citeauthoryear{Kuker \& Rudiger}{2011}]{kuker}Kuker M., Rudiger G., 2011, Astron. Nachr., 332, 933

\bibitem[\protect\citeauthoryear{Kwag \& Kim}{2017}]{kwag}
Kwag, S. G., \& Kim, J. H. 2017, Korean J Anesthesiol., 70(2), 144

\bibitem[\protect\citeauthoryear{Lanza et al.}{2003}]{lanza}Lanza, A. F., Rodono, M., Pagano, I., Barge, P., \& Llebaria, A. 2003, A\&A, 403, 1135

\bibitem[\protect\citeauthoryear{Lanza et al.}{2014}]{lanza2014}
Lanza, A. F.; Das Chagas, M. L. and De Medeiros, J.R. 2014, \aap, 564, A50



\bibitem[\protect\citeauthoryear{Movahed et al.}{2006}]{movahed}
Movahed, M.S., Jafari, G.R., Ghasemi, F., Rahvar, S., \& Reza, M. R. T., 2006. Stat. Mech., 02003

\bibitem[\protect\citeauthoryear{Mandelbrot \& Wallis}{1969a}]{mw1969a}
Mandelbrot, B., \& Wallis, J. R. 1969a, Water Resour. Res., 5, 521

\bibitem[\protect\citeauthoryear{Mandelbrot \& Wallis}{1969b}]{mw1969b}
Mandelbrot, B., \& Wallis, J. R. 1969b, Water Resour. Res., 5, 967

\bibitem[\protect\citeauthoryear{Mathur et al.}{2014}]{mathur2014a}
Mathur, S., Salabert, D., Garcia, R. A., \& Ceillier, T. 2014a, J. Space Weather Space Clim, 4, 15

\bibitem[\protect\citeauthoryear{Mukaka}{2012}]{mukaka}
Mukaka, M. 2012, Malawi Medical Journal, 24, 69

\bibitem[\protect\citeauthoryear{Norouzzadeha, Dullaertc \& Rahmani} {2007}]{Norouzzadeha}
Norouzzadeha, P., Dullaertc, W., \& Rahmani, B. 2007, Physica A, 380, 333

\bibitem[\protect\citeauthoryear{Noyes et al.}{1984}]{noyes}Noyes, R.W., Hartmann, L.W., Baliunas, S.L., Duncan, D.K. \& Vaughan, A .H., 1984, ApJ, 279, 763

\bibitem[\protect\citeauthoryear{Pinsonneault et al.}{2012}]{pen}Pinsonneault, M. H., An, D., Molenda-Zakowicz, J., et al. 2012, ApJS, 199, 30

\bibitem[\protect\citeauthoryear{Press et al.}{2007}]{press}
Press, W. H., Saul A. T., William T. V., \& Brian P. F. 2007, Numerical Recipes in C: The Art of Scientific Computing, Third Edition, Cambridge University Press.


\bibitem[\protect\citeauthoryear{Reinhold et al.}{2013}]{reinhold}Reinhold T., Reiners A., Basri G., 2013, A\&A, 560, A4

\bibitem[\protect\citeauthoryear{Reiners \& Schmitt}{2003}]{Reiners2003}
Reiners, A. \& Schmitt, J.H.M.M. 2003, \aap., 398, 647


\bibitem[\protect\citeauthoryear{Seuront}{2010}]{seuront}
Seuront, L. 2010. Fractals and Multifractals in Ecology and Aquatic Science. CRC
Press, Boca Raton. 

\bibitem[\protect\citeauthoryear{Strassmeier}{2009}]{Strassmeier2009}
Strassmeier, K.G.2009, Cosmic Magnetic Fields: From Planets, to Stars and Galaxies, 259, 363

\bibitem[\protect\citeauthoryear{Suyal, Prasad \& Singh}{2009}]{sps2009}
Suyal, V., Prasad, A., \& Singh, H. P. 2009, Solar Phys., 260, 441

\bibitem[\protect\citeauthoryear{Tanna \& Pathak}{2014}]{tanna}Tanna, H.J., Pathak, K.N., 2014, Astrophys. Space Sci., 350, 47



\bibitem[\protect\citeauthoryear{Tang et al.}{2015}]{tang}
Tang, L., Lv, H., Yang, F., \& Yu, L. 2015, Chaos, Solitons \& Fractals, 81, 117


\bibitem[\protect\citeauthoryear{Teslesca \& Lapenna}{2006}]{telesca2006}
Telesca, L., \& Lapenna V. 2006, Tecnophys., 423, 115

\bibitem[\protect\citeauthoryear{Thompson et al.}{2010}]{thompson}
Thompson, S. E., Christiansen, J. L. , Jenkins, J. M., et al. 2
013, Kepler Data Release 23 Notes (KSCI-19063-001)

\bibitem[\protect\citeauthoryear{Trauth}{2006}]{trauth}Trauth, M. H. 2006, MATLAB Recipes for Earth Sciences Springer, Berlin Heidelberg New York


\bibitem[\protect\citeauthoryear{Van Cleve \& Caldwell}{2009}]{van2}
Van Cleve, J. E. \& Caldwell, D. A. 2009, Kepler Instrument
Handbook, KSCI-19033

\bibitem[\protect\citeauthoryear{Van Cleve et al.}{2010}]{van}Van Cleve, J. E., Jenkins, J. M., Caldwell, D. A., et al. 2010,
Kepler Data Release 5 Notes, KSCI-19045-001

\bibitem[\protect\citeauthoryear{Wengert}{2010}]{wen}
Wengert, C. ``Multifractal  Model  of  Asset  Returns  (MMAR).'' Last  modified  December 12,  2010.  

\end{thebibliography}


\begin{center}
\begin{longtable}{cccccccccccccc}
	\caption{Results of the geometric analysis of the multifractal spectrum for the indexes $A$, $\Delta\alpha$, $\Delta f_{L}(\alpha)$, $\Delta f_{R}(\alpha)$ and $H$ for the Sun and 141 stars with differential rotation, using the SAP4 flux. The values of $P1$	$P2$, $T_{\rm eff}$, $\log g$, $M$, $P_{\rm rot}$, $\beta$	and	$\Delta\Omega$ were extracted from Reinhold et al. (2013). } \label{tab:long} \\ \hline
	
	\hline  
	KIC	&	$P1$	&	$P2$	&	$T_{\rm eff}$	&	$\log g$	&	$M$	&	$P_{\rm rot}$	&	$A$	&	$\Delta\alpha$	&	$\Delta f_{L}(\alpha)$	&	$\Delta f_{R}(\alpha)$	&	$H$	&	$\beta$	&	$\Delta\Omega$	\\
	&	days	&	days	&	K	&	cm/s$^2$	&	$M_{\sun}$	&	days	&		&		&		&		&		&		&	rad/day	\\ \hline 
	\endfirsthead
	
	\multicolumn{14}{c}%
	{{\bfseries \tablename\ \thetable{} -- continued from previous page}} \\
	\hline 
	KIC	&	$P1$	&	$P2$	&	$T_{\rm eff}$	&	$\log g$	&	$M$	&	$P_{\rm rot}$	&	$A$	&	$\Delta\alpha$	&	$\Delta f_{L}(\alpha)$	&	$\Delta f_{R}(\alpha)$	&	$H$	&	$\beta$	&	$\Delta\Omega$	\\ 
	&	days	&	days	&	K	&	cm/s$^2$	&	$M_{\sun}$	&	days	&		&		&		&		&		&		&	rad/day	\\ \hline
	\endhead
	
	\hline \multicolumn{3}{c}{{Continued on next page}} \\ \hline
	\endfoot
	
	\hline \hline
	\endlastfoot
	\hline
Sun	&	24.5	&	33.5	&	5777	&	4.40	&	1.00	&	27.4	&	1.72	&	0.76	&	0.31	&	0.60	&	0.84	&	0.20	&	0.07	\\
1027740	&	26.57	&	20.46	&	5751	&	4.13	&	1.03	&	26.70	&	4.21	&	0.82	&	0.15	&	0.852	&	0.96	&	0.23	&	0.07	\\
1432022	&	30.69	&	23.12	&	5483	&	4.77	&	0.96	&	24.63	&	1.24	&	1.06	&	0.84	&	0.931	&	0.78	&	0.25	&	0.07	\\
2450531	&	22.53	&	18.16	&	5372	&	4.50	&	0.93	&	25.14	&	1.31	&	0.48	&	0.34	&	0.522	&	0.92	&	0.19	&	0.07	\\
2696717	&	22.53	&	17.12	&	5472	&	4.42	&	0.96	&	25.25	&	1.96	&	0.83	&	0.44	&	0.811	&	0.88	&	0.24	&	0.09	\\
2831979	&	25.43	&	20.23	&	5529	&	4.36	&	0.97	&	24.38	&	2.71	&	0.86	&	0.22	&	0.785	&	0.93	&	0.21	&	0.06	\\
2855272	&	22.25	&	28.26	&	5583	&	4.53	&	0.99	&	25.41	&	1.10	&	0.81	&	0.62	&	0.586	&	0.82	&	0.21	&	0.06	\\
3003084	&	23.42	&	28.26	&	5490	&	4.16	&	0.96	&	24.75	&	2.04	&	0.97	&	0.46	&	0.95	&	0.86	&	0.17	&	0.05	\\
3540789	&	19.78	&	25.07	&	5507	&	4.65	&	0.97	&	24.09	&	2.32	&	0.86	&	0.28	&	0.808	&	0.89	&	0.21	&	0.07	\\
3643036	&	12.19	&	10.53	&	5511	&	4.53	&	0.97	&	24.12	&	2.02	&	0.91	&	0.46	&	0.828	&	0.86	&	0.14	&	0.08	\\
3646779	&	23.42	&	16.64	&	5358	&	4.71	&	0.93	&	25.65	&	1.87	&	0.66	&	0.34	&	0.556	&	0.92	&	0.29	&	0.11	\\
4038557	&	11.95	&	10.29	&	5389	&	4.45	&	0.94	&	24.22	&	0.90	&	0.73	&	0.75	&	0.684	&	0.75	&	0.14	&	0.08	\\
4048325	&	26.97	&	34.90	&	5432	&	4.68	&	0.95	&	24.29	&	1.91	&	1.01	&	0.54	&	0.866	&	0.88	&	0.23	&	0.05	\\
4067498	&	24.39	&	18.54	&	5537	&	4.66	&	0.98	&	24.24	&	2.62	&	0.93	&	0.38	&	0.964	&	0.86	&	0.24	&	0.08	\\
4150703	&	23.42	&	18.74	&	5411	&	4.61	&	0.94	&	24.23	&	1.61	&	0.63	&	0.30	&	0.522	&	0.88	&	0.20	&	0.07	\\
4355501	&	28.71	&	20.70	&	5384	&	4.41	&	0.93	&	25.16	&	1.45	&	0.46	&	0.20	&	0.414	&	0.92	&	0.28	&	0.08	\\
4570519	&	24.06	&	29.67	&	5488	&	4.82	&	0.96	&	24.74	&	1.08	&	0.46	&	0.29	&	0.373	&	0.91	&	0.19	&	0.05	\\
4820741	&	24.72	&	20.46	&	5427	&	4.29	&	0.95	&	24.71	&	2.16	&	0.78	&	0.43	&	0.777	&	0.91	&	0.17	&	0.05	\\
5009162	&	23.42	&	16.95	&	5573	&	4.58	&	0.98	&	24.17	&	2.57	&	0.61	&	0.19	&	0.634	&	0.95	&	0.28	&	0.10	\\
5091569	&	32.37	&	24.39	&	5500	&	4.54	&	0.97	&	26.45	&	0.93	&	0.25	&	0.14	&	0.187	&	0.96	&	0.25	&	0.06	\\
5201089	&	28.26	&	22.53	&	5439	&	4.39	&	0.95	&	28.16	&	1.40	&	0.33	&	0.23	&	0.312	&	0.96	&	0.20	&	0.06	\\
5352708	&	25.80	&	19.56	&	5756	&	4.50	&	1.04	&	25.25	&	3.16	&	0.78	&	0.17	&	0.807	&	0.94	&	0.24	&	0.08	\\
5353518	&	23.74	&	17.12	&	5388	&	4.54	&	0.94	&	24.20	&	1.47	&	0.48	&	0.20	&	0.334	&	0.93	&	0.28	&	0.10	\\
5525733	&	19.56	&	24.06	&	5434	&	4.55	&	0.95	&	24.20	&	2.84	&	0.52	&	0.07	&	0.458	&	0.94	&	0.19	&	0.06	\\
5530623	&	20.23	&	17.28	&	5702	&	4.59	&	1.02	&	24.06	&	1.83	&	0.55	&	0.22	&	0.444	&	0.93	&	0.15	&	0.05	\\
5709012	&	30.69	&	22.25	&	5413	&	4.80	&	0.94	&	26.27	&	0.91	&	0.58	&	0.48	&	0.341	&	0.89	&	0.28	&	0.08	\\
5966966	&	25.43	&	20.70	&	5442	&	4.91	&	0.95	&	25.27	&	3.07	&	0.98	&	0.34	&	0.993	&	0.89	&	0.19	&	0.06	\\
6198533	&	25.43	&	19.35	&	5641	&	4.02	&	1.00	&	27.54	&	2.50	&	0.98	&	0.47	&	0.9	&	0.91	&	0.24	&	0.08	\\
6286724	&	30.17	&	22.53	&	5371	&	4.84	&	0.93	&	24.22	&	1.20	&	0.68	&	0.44	&	0.6	&	0.82	&	0.25	&	0.07	\\
6347656	&	14.71	&	18.16	&	5414	&	4.35	&	0.94	&	28.44	&	1.82	&	0.46	&	0.20	&	0.515	&	0.92	&	0.19	&	0.08	\\
6442094	&	28.71	&	21.45	&	5428	&	4.77	&	0.95	&	24.80	&	1.34	&	0.88	&	0.63	&	0.797	&	0.79	&	0.25	&	0.07	\\
6526893	&	28.26	&	20.94	&	5435	&	4.89	&	0.95	&	27.53	&	0.93	&	0.76	&	0.58	&	0.479	&	0.78	&	0.26	&	0.08	\\
6762594	&	24.39	&	19.14	&	5462	&	4.42	&	0.96	&	24.31	&	2.11	&	0.77	&	0.41	&	0.637	&	0.90	&	0.22	&	0.07	\\
6764598	&	26.57	&	20.94	&	5433	&	4.61	&	0.95	&	25.96	&	0.97	&	0.66	&	0.52	&	0.459	&	0.80	&	0.21	&	0.06	\\
6836955	&	28.26	&	22.53	&	5344	&	4.61	&	0.92	&	26.35	&	3.94	&	0.80	&	0.13	&	0.809	&	0.95	&	0.20	&	0.06	\\
6867026	&	26.57	&	18.74	&	5531	&	4.61	&	0.97	&	26.09	&	2.19	&	0.51	&	0.21	&	0.462	&	0.96	&	0.30	&	0.10	\\
6960242	&	15.21	&	10.92	&	5336	&	4.36	&	0.92	&	24.93	&	4.29	&	0.92	&	0.17	&	0.989	&	0.91	&	0.28	&	0.16	\\
7037130	&	27.81	&	23.74	&	5455	&	4.67	&	0.95	&	26.30	&	2.85	&	0.92	&	0.40	&	0.992	&	0.90	&	0.15	&	0.04	\\
7037804	&	24.06	&	19.56	&	5498	&	4.68	&	0.96	&	24.43	&	1.05	&	0.80	&	0.70	&	0.438	&	0.86	&	0.19	&	0.06	\\
7104190	&	25.80	&	20.00	&	5558	&	4.57	&	0.98	&	25.09	&	1.22	&	0.87	&	0.69	&	0.637	&	0.85	&	0.23	&	0.07	\\
7117330	&	12.90	&	10.92	&	5536	&	4.35	&	0.97	&	24.87	&	1.91	&	1.02	&	0.58	&	0.903	&	0.83	&	0.15	&	0.09	\\
7204907	&	23.12	&	17.98	&	5447	&	4.93	&	0.95	&	24.69	&	1.00	&	0.93	&	0.75	&	0.648	&	0.75	&	0.22	&	0.08	\\
7255509	&	21.98	&	18.16	&	5433	&	4.61	&	0.95	&	24.43	&	2.85	&	0.71	&	0.17	&	0.609	&	0.96	&	0.17	&	0.06	\\
7267949	&	37.09	&	26.18	&	5413	&	4.41	&	0.94	&	25.11	&	1.31	&	0.92	&	0.69	&	0.514	&	0.88	&	0.29	&	0.07	\\
7430659	&	29.18	&	37.09	&	5566	&	4.28	&	0.98	&	28.39	&	4.37	&	0.90	&	0.18	&	0.892	&	0.96	&	0.21	&	0.05	\\
7517798	&	26.57	&	20.23	&	5415	&	4.70	&	0.94	&	24.77	&	1.64	&	1.01	&	0.65	&	0.829	&	0.85	&	0.24	&	0.07	\\
7586869	&	28.71	&	22.53	&	5422	&	4.43	&	0.94	&	26.57	&	2.89	&	0.52	&	0.13	&	0.583	&	0.96	&	0.22	&	0.06	\\
7750731	&	30.17	&	23.12	&	5491	&	4.57	&	0.96	&	27.24	&	1.19	&	1.01	&	0.81	&	0.708	&	0.82	&	0.23	&	0.06	\\
7798090	&	27.81	&	23.12	&	5558	&	4.43	&	0.98	&	26.84	&	6.72	&	0.90	&	0.05	&	0.972	&	0.98	&	0.17	&	0.05	\\
7808399	&	13.69	&	9.62	&	5667	&	4.51	&	1.01	&	28.73	&	1.11	&	0.62	&	0.48	&	0.391	&	0.89	&	0.30	&	0.19	\\
7879315	&	24.06	&	18.94	&	5472	&	4.31	&	0.96	&	24.35	&	3.83	&	1.04	&	0.30	&	0.986	&	0.95	&	0.21	&	0.07	\\
7938914	&	25.80	&	18.16	&	5568	&	4.51	&	0.98	&	24.47	&	3.10	&	0.95	&	0.31	&	0.997	&	0.91	&	0.30	&	0.10	\\
8022329	&	34.23	&	25.07	&	5448	&	4.46	&	0.95	&	28.17	&	1.13	&	0.93	&	0.75	&	0.619	&	0.79	&	0.27	&	0.07	\\
8029079	&	26.97	&	19.14	&	5451	&	4.36	&	0.95	&	28.03	&	1.92	&	0.88	&	0.48	&	0.829	&	0.90	&	0.29	&	0.10	\\
8074818	&	26.97	&	20.23	&	5403	&	4.56	&	0.94	&	24.17	&	3.39	&	0.76	&	0.14	&	0.747	&	0.93	&	0.25	&	0.08	\\
8093179	&	28.71	&	20.94	&	5616	&	4.52	&	1.00	&	28.26	&	0.91	&	0.68	&	0.58	&	0.572	&	0.80	&	0.27	&	0.08	\\
8094374	&	30.17	&	24.06	&	5416	&	4.65	&	0.94	&	27.35	&	1.21	&	0.71	&	0.57	&	0.525	&	0.84	&	0.20	&	0.05	\\
8095149	&	26.18	&	21.19	&	5497	&	4.84	&	0.96	&	24.22	&	1.46	&	1.09	&	0.74	&	0.819	&	0.81	&	0.19	&	0.06	\\
8145579	&	23.74	&	19.56	&	5525	&	4.43	&	0.97	&	25.04	&	3.63	&	0.70	&	0.09	&	0.724	&	0.95	&	0.18	&	0.06	\\
8153705	&	23.42	&	19.35	&	5407	&	4.61	&	0.94	&	24.31	&	1.06	&	0.68	&	0.57	&	0.433	&	0.84	&	0.17	&	0.06	\\
8212665	&	32.37	&	23.42	&	5530	&	4.42	&	0.97	&	30.99	&	3.52	&	0.76	&	0.15	&	0.798	&	0.95	&	0.28	&	0.07	\\
8231309	&	26.18	&	20.70	&	5528	&	4.47	&	0.97	&	24.47	&	1.95	&	0.83	&	0.47	&	0.778	&	0.87	&	0.21	&	0.06	\\
8282808	&	26.97	&	20.94	&	5541	&	4.45	&	0.98	&	26.08	&	2.31	&	0.65	&	0.33	&	0.742	&	0.95	&	0.22	&	0.07	\\
8351578	&	26.18	&	21.19	&	5658	&	4.40	&	1.01	&	26.06	&	3.22	&	0.65	&	0.14	&	0.556	&	0.97	&	0.19	&	0.06	\\
8360859	&	19.78	&	16.48	&	5554	&	4.71	&	0.98	&	24.52	&	0.67	&	0.56	&	0.58	&	0.364	&	0.81	&	0.17	&	0.06	\\
8429890	&	30.17	&	24.39	&	5423	&	4.71	&	0.94	&	27.32	&	0.02	&	0.33	&	0.78	&	0.525	&	0.85	&	0.19	&	0.05	\\
8491487	&	20.46	&	17.98	&	5511	&	4.68	&	0.97	&	25.23	&	1.13	&	0.99	&	0.66	&	0.702	&	0.72	&	0.12	&	0.04	\\
8493268	&	26.97	&	22.53	&	5422	&	4.50	&	0.94	&	27.27	&	1.55	&	0.98	&	0.55	&	0.724	&	0.84	&	0.17	&	0.05	\\
8495770	&	25.07	&	18.94	&	5489	&	4.56	&	0.96	&	25.63	&	2.18	&	0.64	&	0.21	&	0.562	&	0.93	&	0.25	&	0.08	\\
8553548	&	24.39	&	19.35	&	5575	&	4.91	&	0.99	&	24.38	&	1.61	&	1.00	&	0.55	&	0.533	&	0.86	&	0.21	&	0.07	\\
8556908	&	24.39	&	19.35	&	5518	&	4.68	&	0.97	&	24.67	&	4.21	&	0.97	&	0.19	&	0.998	&	0.94	&	0.21	&	0.07	\\
8616755	&	34.23	&	25.43	&	5512	&	4.55	&	0.97	&	27.93	&	4.05	&	0.81	&	0.18	&	0.903	&	0.91	&	0.26	&	0.06	\\
8620514	&	27.81	&	34.90	&	5532	&	4.69	&	0.97	&	28.05	&	1.32	&	0.94	&	0.74	&	0.782	&	0.85	&	0.20	&	0.05	\\
8771390	&	25.43	&	20.00	&	5417	&	4.22	&	0.94	&	25.21	&	1.98	&	0.40	&	0.16	&	0.523	&	0.93	&	0.21	&	0.07	\\
8871159	&	26.18	&	21.98	&	5336	&	4.48	&	0.92	&	25.18	&	3.38	&	0.72	&	0.18	&	0.832	&	0.95	&	0.16	&	0.05	\\
8957265	&	30.17	&	24.72	&	5434	&	4.39	&	0.95	&	25.91	&	2.49	&	0.86	&	0.49	&	0.901	&	0.94	&	0.18	&	0.05	\\
8962193	&	25.80	&	19.78	&	5383	&	4.72	&	0.93	&	27.49	&	3.07	&	0.55	&	0.17	&	0.711	&	0.96	&	0.23	&	0.07	\\
8971874	&	25.43	&	19.78	&	5478	&	4.49	&	0.96	&	24.08	&	2.37	&	0.51	&	0.11	&	0.472	&	0.94	&	0.22	&	0.07	\\
9028802	&	26.18	&	19.78	&	5552	&	4.41	&	0.98	&	25.66	&	1.92	&	0.81	&	0.51	&	0.898	&	0.89	&	0.24	&	0.08	\\
9071286	&	13.28	&	11.79	&	5537	&	4.69	&	0.98	&	26.07	&	2.24	&	0.58	&	0.20	&	0.594	&	0.91	&	0.11	&	0.06	\\
9076541	&	21.71	&	15.62	&	5493	&	4.62	&	0.96	&	24.70	&	1.94	&	0.91	&	0.33	&	0.645	&	0.88	&	0.28	&	0.11	\\
9279550	&	23.12	&	16.48	&	5429	&	4.62	&	0.95	&	25.54	&	3.15	&	0.71	&	0.13	&	0.653	&	0.92	&	0.29	&	0.11	\\
9407118	&	25.07	&	31.23	&	5548	&	4.71	&	0.98	&	24.18	&	4.51	&	0.65	&	0.16	&	0.945	&	0.97	&	0.20	&	0.05	\\
9414450	&	25.07	&	19.78	&	5643	&	4.71	&	1.00	&	25.37	&	1.65	&	0.71	&	0.47	&	0.587	&	0.90	&	0.21	&	0.07	\\
9447022	&	13.09	&	11.71	&	5459	&	4.57	&	0.95	&	27.64	&	1.82	&	0.57	&	0.23	&	0.37	&	0.93	&	0.11	&	0.06	\\
9449447	&	29.67	&	23.12	&	5462	&	4.43	&	0.96	&	27.19	&	3.91	&	0.74	&	0.14	&	0.818	&	0.96	&	0.22	&	0.06	\\
9581214	&	28.26	&	21.98	&	5378	&	4.89	&	0.93	&	27.50	&	2.70	&	0.58	&	0.25	&	0.64	&	0.94	&	0.22	&	0.06	\\
9657969	&	32.37	&	24.72	&	5478	&	4.75	&	0.96	&	26.54	&	2.00	&	0.95	&	0.48	&	0.99	&	0.86	&	0.24	&	0.06	\\
9700186	&	27.81	&	22.53	&	5497	&	4.09	&	0.96	&	26.06	&	2.80	&	0.84	&	0.18	&	0.838	&	0.96	&	0.19	&	0.05	\\
9813448	&	25.07	&	18.35	&	5489	&	4.16	&	0.96	&	24.01	&	4.20	&	0.75	&	0.13	&	0.782	&	0.95	&	0.27	&	0.09	\\
9825207	&	26.18	&	20.46	&	5342	&	4.58	&	0.92	&	25.35	&	2.96	&	0.58	&	0.14	&	0.593	&	0.90	&	0.22	&	0.07	\\
9840009	&	29.18	&	22.82	&	5494	&	4.65	&	0.96	&	27.69	&	2.02	&	0.67	&	0.19	&	0.512	&	0.94	&	0.22	&	0.06	\\
9934371	&	24.39	&	20.23	&	5545	&	4.06	&	0.98	&	26.92	&	5.80	&	0.64	&	0.05	&	0.746	&	0.96	&	0.17	&	0.05	\\
9963105	&	27.39	&	21.71	&	5552	&	4.38	&	0.98	&	28.09	&	0.57	&	0.26	&	0.33	&	0.253	&	0.93	&	0.21	&	0.06	\\
9964492	&	22.82	&	18.74	&	5605	&	4.40	&	0.99	&	24.10	&	1.82	&	0.41	&	0.19	&	0.435	&	0.94	&	0.18	&	0.06	\\
9996105	&	27.81	&	22.82	&	5585	&	3.99	&	0.99	&	28.68	&	2.86	&	0.69	&	0.18	&	0.615	&	0.97	&	0.18	&	0.05	\\
10081120	&	24.72	&	20.00	&	5492	&	4.76	&	0.96	&	24.97	&	2.09	&	0.61	&	0.16	&	0.461	&	0.95	&	0.19	&	0.06	\\
10125100	&	31.23	&	25.43	&	5417	&	4.61	&	0.94	&	26.54	&	2.49	&	0.57	&	0.13	&	0.585	&	0.95	&	0.19	&	0.05	\\
10189540	&	25.07	&	20.23	&	5459	&	4.54	&	0.95	&	25.51	&	3.58	&	0.64	&	0.12	&	0.572	&	0.97	&	0.19	&	0.06	\\
10192060	&	28.26	&	21.45	&	5409	&	4.41	&	0.94	&	27.86	&	3.30	&	0.63	&	0.19	&	0.727	&	0.96	&	0.24	&	0.07	\\
10196130	&	13.91	&	11.13	&	5507	&	4.53	&	0.97	&	27.68	&	3.30	&	0.63	&	0.15	&	0.654	&	0.91	&	0.20	&	0.11	\\
10220780	&	24.72	&	20.23	&	5420	&	4.72	&	0.94	&	24.07	&	2.31	&	0.68	&	0.28	&	0.572	&	0.92	&	0.18	&	0.06	\\
10287800	&	24.72	&	20.23	&	5595	&	4.51	&	0.99	&	24.04	&	4.12	&	0.56	&	0.10	&	0.772	&	0.96	&	0.18	&	0.06	\\
10387740	&	29.18	&	22.53	&	5448	&	4.56	&	0.95	&	30.95	&	2.52	&	0.58	&	0.23	&	0.722	&	0.94	&	0.23	&	0.06	\\
10454270	&	12.54	&	11.34	&	5379	&	4.72	&	0.93	&	24.53	&	3.81	&	0.92	&	0.18	&	0.966	&	0.90	&	0.10	&	0.05	\\
10483450	&	25.43	&	19.78	&	5551	&	4.73	&	0.98	&	24.86	&	3.44	&	0.70	&	0.13	&	0.64	&	0.93	&	0.22	&	0.07	\\
10592740	&	21.71	&	18.16	&	5484	&	4.50	&	0.96	&	26.95	&	1.87	&	0.56	&	0.34	&	0.559	&	0.91	&	0.16	&	0.06	\\
10610300	&	32.97	&	25.07	&	5528	&	4.17	&	0.97	&	30.31	&	4.16	&	0.89	&	0.13	&	0.779	&	0.97	&	0.24	&	0.06	\\
10651440	&	20.70	&	25.07	&	5441	&	4.85	&	0.95	&	24.42	&	3.28	&	0.68	&	0.14	&	0.782	&	0.92	&	0.17	&	0.05	\\
10652090	&	14.02	&	15.62	&	5421	&	4.45	&	0.94	&	28.08	&	1.53	&	0.75	&	0.43	&	0.718	&	0.84	&	0.10	&	0.05	\\
10658860	&	24.39	&	20.00	&	5500	&	4.49	&	0.97	&	25.12	&	4.27	&	0.79	&	0.15	&	0.978	&	0.94	&	0.18	&	0.06	\\
10670950	&	24.72	&	20.23	&	5448	&	4.13	&	0.95	&	27.53	&	4.50	&	0.78	&	0.11	&	0.743	&	0.97	&	0.18	&	0.06	\\
10724980	&	24.06	&	17.45	&	5732	&	4.62	&	1.03	&	24.24	&	1.90	&	0.52	&	0.20	&	0.488	&	0.92	&	0.27	&	0.10	\\
10748260	&	22.53	&	18.74	&	5469	&	4.69	&	0.96	&	24.07	&	5.49	&	0.83	&	0.09	&	0.961	&	0.96	&	0.17	&	0.06	\\
10864340	&	30.69	&	22.82	&	5530	&	4.56	&	0.97	&	28.39	&	2.22	&	0.43	&	0.13	&	0.495	&	0.96	&	0.26	&	0.07	\\
10972390	&	25.07	&	30.69	&	5552	&	4.27	&	0.98	&	24.31	&	3.36	&	0.99	&	0.19	&	0.992	&	0.91	&	0.18	&	0.05	\\
10972620	&	13.49	&	17.12	&	5608	&	4.48	&	0.99	&	26.42	&	2.22	&	0.58	&	0.24	&	0.565	&	0.92	&	0.21	&	0.10	\\
11097790	&	25.80	&	19.35	&	5573	&	4.53	&	0.98	&	24.73	&	2.25	&	0.42	&	0.21	&	0.564	&	0.94	&	0.25	&	0.08	\\
11147150	&	30.69	&	24.39	&	5413	&	4.87	&	0.94	&	25.12	&	1.68	&	0.31	&	0.14	&	0.38	&	0.95	&	0.21	&	0.05	\\
11245710	&	26.57	&	20.46	&	5382	&	4.54	&	0.93	&	25.59	&	3.49	&	0.82	&	0.14	&	0.771	&	0.96	&	0.23	&	0.07	\\
11305180	&	27.39	&	20.70	&	5530	&	4.59	&	0.97	&	25.59	&	3.44	&	0.81	&	0.18	&	0.865	&	0.96	&	0.24	&	0.07	\\
11413690	&	28.26	&	22.25	&	5626	&	4.77	&	1.00	&	24.24	&	0.68	&	0.21	&	0.11	&	0.171	&	0.92	&	0.21	&	0.06	\\
11446080	&	29.18	&	24.72	&	5408	&	4.64	&	0.94	&	27.17	&	2.13	&	0.77	&	0.23	&	0.637	&	0.93	&	0.15	&	0.04	\\
11454110	&	14.13	&	17.45	&	5488	&	4.82	&	0.96	&	27.44	&	1.38	&	0.98	&	0.71	&	0.671	&	0.79	&	0.19	&	0.08	\\
11455230	&	26.18	&	20.46	&	5577	&	4.51	&	0.99	&	25.60	&	1.42	&	0.16	&	0.09	&	0.292	&	0.97	&	0.22	&	0.07	\\
11462110	&	12.45	&	14.96	&	5459	&	4.43	&	0.95	&	24.53	&	3.69	&	0.88	&	0.16	&	0.985	&	0.91	&	0.17	&	0.08	\\
11509110	&	25.80	&	21.45	&	5544	&	4.79	&	0.98	&	29.76	&	2.02	&	0.58	&	0.29	&	0.629	&	0.94	&	0.17	&	0.05	\\
11548990	&	12.81	&	15.35	&	5492	&	4.53	&	0.96	&	25.31	&	1.86	&	0.57	&	0.19	&	0.451	&	0.88	&	0.17	&	0.08	\\
11603050	&	23.12	&	19.14	&	5675	&	4.60	&	1.01	&	27.60	&	2.45	&	0.75	&	0.19	&	0.688	&	0.91	&	0.17	&	0.06	\\
11764150	&	27.39	&	19.35	&	5423	&	4.61	&	0.94	&	26.43	&	2.94	&	0.86	&	0.34	&	0.955	&	0.91	&	0.29	&	0.10	\\
11818770	&	28.26	&	21.98	&	5605	&	4.62	&	0.99	&	27.02	&	2.70	&	0.85	&	0.36	&	0.886	&	0.92	&	0.22	&	0.06	\\
11854160	&	25.43	&	20.70	&	5502	&	4.39	&	0.97	&	25.12	&	0.02	&	0.30	&	0.74	&	0.162	&	0.89	&	0.19	&	0.06	\\
11909840	&	13.49	&	12.03	&	5464	&	4.53	&	0.96	&	28.06	&	2.45	&	0.64	&	0.15	&	0.665	&	0.89	&	0.11	&	0.06	\\
11911790	&	22.53	&	15.89	&	5406	&	4.73	&	0.94	&	24.07	&	1.92	&	0.72	&	0.36	&	0.571	&	0.89	&	0.30	&	0.12	\\
11955130	&	25.07	&	20.23	&	5588	&	4.82	&	0.99	&	24.45	&	2.11	&	0.80	&	0.41	&	0.842	&	0.89	&	0.19	&	0.06	\\
11970930	&	25.07	&	19.35	&	5397	&	4.48	&	0.94	&	27.75	&	2.13	&	0.67	&	0.23	&	0.651	&	0.94	&	0.23	&	0.07	\\
11971380	&	26.97	&	21.98	&	5575	&	4.50	&	0.99	&	27.23	&	2.05	&	0.52	&	0.21	&	0.409	&	0.95	&	0.19	&	0.05	\\
12023900	&	23.74	&	18.94	&	5552	&	4.55	&	0.98	&	26.27	&	3.20	&	0.62	&	0.14	&	0.628	&	0.94	&	0.20	&	0.07	\\
12052640	&	26.97	&	21.98	&	5457	&	4.57	&	0.95	&	24.17	&	2.05	&	0.75	&	0.20	&	0.546	&	0.91	&	0.19	&	0.05	\\
12218610	&	26.57	&	21.45	&	5494	&	4.66	&	0.96	&	24.59	&	4.29	&	0.81	&	0.16	&	0.944	&	0.96	&	0.19	&	0.06	\\
12256700	&	26.97	&	20.70	&	5434	&	4.61	&	0.95	&	24.40	&	2.52	&	1.03	&	0.34	&	0.898	&	0.87	&	0.23	&	0.07	\\
12400710	&	26.57	&	20.00	&	5537	&	4.59	&	0.98	&	25.74	&	2.15	&	0.87	&	0.41	&	0.636	&	0.91	&	0.25	&	0.08	\\
	
\end{longtable}
\end{center}

\end{document}